\documentclass[conference]{IEEEtran}

\usepackage{xcolor}
\usepackage{graphicx}
\usepackage{amssymb}

\newcommand{\cmarkg}{\textcolor{mygreen}{\cmark}}
\newcommand{\xmarkr}{\textcolor{myred}{\xmark}}

\newcommand{\ms}[1]{\mathsmaller{#1}}
\newcommand{\B}{\mathbf{B}}
\newcommand{\vc}[1]{\mathbf{#1}}

\newcommand{\trans}{\ms{T}}
\newcommand{\Um}{\mathbf{U}}
\newcommand{\acro}{\textsc{casad}}
\newcommand{\stacro}{\textsc{Casad}}

\definecolor{myred}{rgb}{0.7,0,0}
\definecolor{mygreen}{rgb}{0,0.7,0}

\usepackage{amsmath}
\usepackage{relsize}
\usepackage{amsfonts}
\usepackage[font=small,skip=1pt]{caption}

\usepackage{microtype} %improves the layout to get more space
\usepackage[numbers,sort,compress]{natbib}
\usepackage{multirow}
\usepackage[font=small,skip=0pt]{subcaption}

\newcommand{\quotes}[1]{``#1''}

\usepackage{amssymb}% http://ctan.org/pkg/amssymb
\usepackage{pifont}% http://ctan.org/pkg/pifont
\newcommand{\cmark}{\ding{51}}%
\newcommand{\xmark}{\ding{55}}%

\begin{document}

%
% The "title" command has an optional parameter, allowing the author to define a "short title" to be used in page headers.
\title{CASAD: CAN-Aware Stealthy-Attack Detection for In-Vehicle Networks}

%
% The "author" command and its associated commands are used to define the authors and their affiliations.
% Of note is the shared affiliation of the first two authors, and the "authornote" and "authornotemark" commands
% used to denote shared contribution to the research.
\author{
\IEEEauthorblockN{Nasser Nowdehi\IEEEauthorrefmark{1},
Wissam Aoudi\IEEEauthorrefmark{2},
Magnus Almgren\IEEEauthorrefmark{2},~and
Tomas Olovsson\IEEEauthorrefmark{2}}
\IEEEauthorblockA{\IEEEauthorrefmark{1}Volvo Cars Corporation,
Gothenburg, Sweden, nasser.nowdehi@volvocars.com}
\IEEEauthorblockA{\IEEEauthorrefmark{2}Chalmers University of Technology, Gothenburg, Sweden}
\IEEEauthorblockA{\IEEEauthorrefmark{2}\{wissam.aoudi,magnus.almgren,tomas.olovsson\}@chalmers.se}
}

\iffalse
\author{Nasser~Nowdehi,
        Wissam~Aoudi,
        Magnus~Almgren,
        and~Tomas~Olovsson
\thanks{Nasser Nowdehi is with Volvo Cars Corporation, Gothenburg, Sweden (e-mail:~nasser.nowdehi@volvocars.com)}%
\thanks{Wissam~Aoudi,~Magnus~Almgren,~and~Tomas~Olovsson are with the Department of Computer Science and Engineering, Chalmers University of Technology, Gothenburg,
Sweden (e-mail:~\{wissam.aoudi,magnus.almgren,tomas.olovsson\}@chalmers.se)}}
\fi
\iffalse
\author{Nasser Nowdehi}
\affiliation{%
  \institution{Volvo Cars Corporation}
  \city{Gothenburg}
  \country{Sweden}}
\email{nasser.nowdehi@volvocars.com}

\author{Wissam Aoudi}
\affiliation{%
  \institution{Chalmers University of Technology}
  \city{Gothenburg}
  \country{Sweden}}
\email{wissam.aoudi@chalmers.se}

\author{Magnus Almgren}
\affiliation{%
 \institution{Chalmers University of Technology}
  \city{Gothenburg}
  \country{Sweden}}
\email{magnus.almgren@chalmers.se}

\author{Tomas Olovsson}
\affiliation{%
  \institution{Chalmers University of Technology}
  \city{Gothenburg}
  \country{Sweden}}
\email{tomas.olovsson@chalmers.se}
\fi
% Use the following at camera-ready time to suppress page numbers.
% Comment it out when you first submit the paper for review.
%\thispagestyle{empty}
\maketitle
\IEEEpeerreviewmaketitle
\begin{abstract}
%\subsection*{Abstract}
Nowadays, vehicles have complex in-vehicle networks (IVNs) with millions of lines of code controlling almost every function in the vehicle including safety-critical functions. It has recently been shown that IVNs are becoming increasingly vulnerable to cyber-attacks capable of taking control of vehicles, thereby threatening the safety of the passengers. Several countermeasures have been proposed in the literature in response to the arising threats, however, hurdle requirements imposed by the industry is hindering their adoption in practice. In particular, detecting attacks on IVNs is challenged by strict resource constraints and utterly complex communication patterns that vary even for vehicles of the same model. In addition, existing solutions suffer from two main drawbacks. First, they depend on the underlying vehicle configuration, and second, they are incapable of detecting certain attacks of a stealthy nature.

In this paper, we propose~\acro{}, a CAN-Aware Stealthy-Attack Detection mechanism that does not abide by the strict specifications predefined for every vehicle model and addresses key real-world deployability challenges. Our fast, lightweight, and system-agnostic approach learns the normal behavior of IVN dynamics from historical data and detects deviations by continuously monitoring IVN traffic. We demonstrate the effectiveness of \acro{} by conducting various experiments on a CAN bus prototype, a 2018 Volvo XC60, and publicly available data from two real vehicles. Our approach is experimentally shown to be effective against different attack scenarios, including the prompt detection of stealthy attacks, and has considerable potential applicability to real vehicles.

%we introduce a novel type of attack that is stealthy, feasible, and has potential to cause serious impact on IVNs. Then,

%, that overcomes the limitations of existing techniques by being capable of detecting stealthy attacks, while not abiding

%Over the past decade there has been a rapid increase in the number of attacks on in-vehicle networks where CAN bus has been the primary target of the attacks. With CAN being the most prevalent protocol used for safety critical applications in vehicles, designing Intrusion Detection Systems (IDS) for CAN communications has become a major area of interest within the field of automotive security. A large number of studies are centered around the idea of monitoring CAN traffic for unlikely deviations in the periodicity of messages, while other studies focus on measuring and utilizing deviations in low-level physical properties of ECUs to identify the attacker. Despite being capable of detecting attacks that cause such kind of deviations, the state-of-the-art IDSs that use
\end{abstract}

\section{Introduction}
\label{sec:introduction}
In-vehicle security has recently attracted notable attention as real-world attacks have demonstrated that it is possible to remotely control vehicles and compromise safety-critical functions via, for example, the Internet-enabled multimedia system, thereby threatening the safety of the passengers~\cite{wright2011hacking,checkoway2011comprehensive, zhao2013challenges, studnia2013survey, kleberger2011security}. 

As early as 2010 and 2011, a group of researchers~\cite{checkoway2011comprehensive, koscher2010experimental} demonstrated two unprecedented remote attacks on a General Motors 2009 Chevrolet Impala allowing them to physically control the vehicle. In the first attack, they
% implemented a Trojan Horse application on a smartphone to 
compromised the vehicle's media player system by exploiting a vulnerability in its Bluetooth stack implementation, while in the second attack, a vulnerability in the telematic unit connected to the safety-critical network was exploited to send arbitrary brake messages. A few years later,~\citet{miller2015remote} compromised a Jeep Cherokee through a vulnerability in its Internet-enabled multimedia system, allowing them to control the steering, the brakes, and the acceleration of the vehicle. Rather alarmingly, they showed how this exploit could be implemented as a worm to quickly compromise approximately 1.4 million vulnerable vehicles. The trend of cyberattacks against vehicles has continued to date with more major brands, such as Tesla and BMW, being remotely compromised~\cite{keen2016tesla,keen2018bmw}. Although some automakers seem to have reacted to certain breaches by issuing over-the-air updates to patch exploited vulnerabilities, it has been shown that the introduced security protection mechanisms can be bypassed~\cite{keen2017tesla}.
%Although the media player device is separated from the safety-critical components via a gateway, they managed to reprogram the gateway device
%through one of its interfaces which they had access to, 
%and send messages to control the brakes. 

%Quite recently, members of Tencent’s Keen Security Lab set out a series of experiments to audit the security of most recent vehicles~\cite{keen2016tesla,keen2017tesla,keen2018bmw}. By exploiting known vulnerabilities in the built-in web browser of 
%running on the CID (i.e. touch screen information display) of 
%a 2016 Tesla Model S, they managed to compromise its CID (the touch screen information display) and subsequently reprogram its gateway node and send messages to engage the brakes~\cite{keen2016tesla}. As an incident response to this breach, Tesla issued over-the-air updates to patch the exploited vulnerability by adding software authentication to the gateway. However, it was later shown how the introduced protection mechanism could be bypassed and the gateway could still be reprogrammed~\cite{keen2017tesla}. The same team of researchers were able to compromise the telematic unit of a 2018 BMW i3 by sending a series of SMS messages, allowing them to send diagnostic commands through its gateway device~\cite{keen2018bmw}. 

%These attacks have raised concerns about security in vehicles and some countries have already enacted legislation to mandate security protections, including intrusion detection and prevention systems, for vehicles \cite{self2017senate, spy2017senate}. 
The increasing susceptibility of vehicles to cyberattacks is in large part due to connectivity to the Internet, lack of secure network partitioning that ensures separation of safety-related domains from the rest of the network, and lack of measures to verify the integrity and authenticity of Electronic Control Unit (ECU) software and communications. In addition, the communication architectures currently used in In-Vehicle Networks (IVNs) were mainly designed with no security in mind. In particular, the Controller Area Network~(CAN), by far the most prevailing bus technology in IVNs, is inherently insecure and lacks the necessary means of protecting against message tampering and spoofing attacks. For instance, CAN messages are broadcast and carry no information about the sender and thus can be easily spoofed. 

To combat the emerging security threats to IVNs, several defensive measures have been proposed in the literature, which can be broadly categorized as \emph{message authentication schemes} and \emph{Intrusion Detection Systems} (IDS). The maximum payload size of a CAN message being only 8 bytes makes appending a cryptographically secure Message Authentication Code (MAC) to the available space a challenge. Several approaches have been proposed to resolve this limitation~\cite{schweppe2011car2x, woo2014practical, herrewege2011canauth,groza2012libra, wang2014vecure, nurnberger2016vatican, weisglass2016authentication}, however,  
%several message authentication schemes have been proposed, some of which suggest using a 32-bit (or less) MAC~\cite{schweppe2011car2x, hartkopp2012macan, woo2014practical, herrewege2011canauth}, while others choose to carry the MAC in one or more separate CAN messages~\cite{groza2012libra, wang2014vecure, nurnberger2016vatican, weisglass2016authentication}. Given that a MAC should have at least 64 bits to provide sufficient protection against guessing attacks,\footnote{According to the US National Institute of Standards and Technology (NIST)~\cite{nist2005cmac}.} 
the proposed schemes hardly make it to the industry due to the tight resource constraints in automotive systems, with backward compatibility and acceptable overhead expectations being the biggest adoption hurdles~\cite{nowdehi2017canauthentication}. 

Intrusion detection systems, on the other hand, are designed to passively monitor IVN traffic for anomalies without imposing computational overhead on in-vehicle communication. As such, they meet the challenging resource constraints and strict real-time requirements of IVNs. In recent years, there have been several attempts to design and develop intrusion detection systems for IVNs~\cite{muter2010structured, muter2011entropy, song2016timeinterval, kang2016intrusion, cho2016fingerprinting}. Due to the long life-span (decades) of vehicles and the difficulty to maintain regular updates, anomaly-based detection has been considered to be more viable than signature-based approaches~\cite{muter2010structured}. 

Most previous studies on anomaly-based attack detection leverage the high regularity of the timing behavior of IVN messages to detect malicious traffic by monitoring for unlikely changes in their periodicity. Other studies leverage the subtle, yet distinctive, differences in the physical properties of ECUs to detect intruders and identify compromised ECUs~\cite{murvay2014source,choi2016identifying,cho2016fingerprinting,Kneib2018}. Since a considerable portion of IVN messages are transmitted periodically, and CAN messages are inherently associated with unique low-level physical ECU properties, existing approaches are, by and large, capable of detecting attacks that cause such kinds of deviation. State-of-the-art solutions, however, fall short on two main fronts. 

First, there have been no noteworthy attempts to detect \emph{stealthy attacks} that do not cause drastic changes in the IVN dynamics. Among the most common types of attacks on IVNs considered in the literature (see Section~\ref{subsec:attackScenarios}), the \emph{masquerade} attack, where the adversary injects attack messages from a compromised ECU while simultaneously muting the intended ECU, is often branded as a stealthy attack. However, in Section~\ref{sec:attackTaxonomy}, we show that the masquerade attack does cause changes in the overall behavior of the CAN traffic and is hardly stealthy. This paper introduces a novel, truly stealthy attack, in which the adversary \emph{reprograms} the intended ECU and directly manipulates the payloads of its messages without affecting the IVN characteristics. We argue that this stealthy attack is feasible, and potentially as severe as the other existing attacks. We find no clear evidence in the literature that existing techniques are capable of detecting this type of attacks. 

Second, in most cases, prior knowledge about the underlying IVN traffic (frame ID, transmission frequency, etc.) and ECU configurations is needed. This makes the existing solutions \emph{dependent on the underlying system specifications}, which are typically proprietary and may vary in vehicles of the same model and year produced by the same OEM, let alone in vehicles of different brands.

In this work, inspired by a recently proposed specification-agnostic method (PASAD) for detecting attacks on industrial control systems (ICSs)~\cite{aoudi2018pasad}, we propose \textbf{\acro{}}, a \textbf{C}AN-\textbf{A}ware \textbf{S}tealthy-\textbf{A}ttack \textbf{D}etection mechanism for IVNs. As IVNs differ considerably from ICSs, we explain in Section~\ref{sec:methodology} how we leverage PASAD's underlying theory to develop \acro{} for the IVN domain.
%We recognize that IVNs are conceptually similar to process-level networks in ICS in the sense that they both operate cyber-physical systems that combine computational elements with physical elements (e.g., controllers, sensors, and actuators) to control physical processes. Furthermore, both IVNs and ICS have strict predefined specifications, real-time requirements, and resource constraints; and both exhibit highly regular dynamics. However, due to distinctive differences between the two domains, a direct application of PASAD to IVNs is infeasible. 

\stacro{} employs an exploratory time-series analysis technique to capture the \emph{deterministic behavior} of the IVN dynamics by processing CAN traffic at the payload level. The method works by first identifying a mathematical representation of the normal behavior of IVNs and subsequently monitoring in real time for attack-indicating changes in the payload structure. In addition to being inherently capable of detecting stealthy attacks on IVNs by detecting slight variations in the monitored signal, \acro{} requires no prior knowledge of the mechanism generating the CAN traffic. As such, the proposed technique overcomes the discussed challenges pertaining to the deployability of such solutions in practice.

More specifically, this paper makes the following contributions: \textit{(i)} Inspired by an existing technique from a different domain, we introduce \acro{}; a fast, lightweight, and specification-agnostic attack-detection mechanism for IVNs that goes a long way toward overcoming adoption hurdles imposed by the industry; \textit{(ii)} We demonstrate the effectiveness of our proposed approach by conducting extensive experiments including performing attacks that we designed to serve as real-world scenarios on a 2018 Volvo XC60; \textit{(iii)} We introduce a novel, stealthy, and overlooked type of attack in which the adversary is capable of manipulating message payloads directly at the target system while remaining under the radar; \textit{(iv)} We show that by monitoring CAN traffic in a way that the entire stream of CAN message payloads is treated as a single signal, we require no comprehension of the actual encoded signals and the underlying vehicle specifications that are typically proprietary. As such, our approach is applicable to vehicles of different brands and configurations; \textit{(v)} We show that, by identifying malicious manipulations directly at the payload level, \acro{} is capable of detecting strategic adversaries who ensure that message frequencies and low-level ECU configurations remain intact under the attack.

After presenting background material and related work in Section~\ref{sec:background}, we underline the vulnerability of IVNs and describe attack scenarios in Section~\ref{sec:attackTaxonomy}.~Section~\ref{sec:methodology} presents our specification-agnostic detection methodology for IVNs, which we evaluate in Section~\ref{sec:experiment}. Finally, we conclude this work in Section~\ref{sec:conclusion}.
\section{Background}
\label{sec:background}
The CAN protocol is the primary target of state-of-the-art IVN intrusion detection systems. This section lays down the principles of CAN communication followed by a discussion of related work. 
\subsection{CAN Communication}
\label{subsec:can}
\begin{figure}[!t]
    \centering
	\includegraphics[width=\columnwidth]{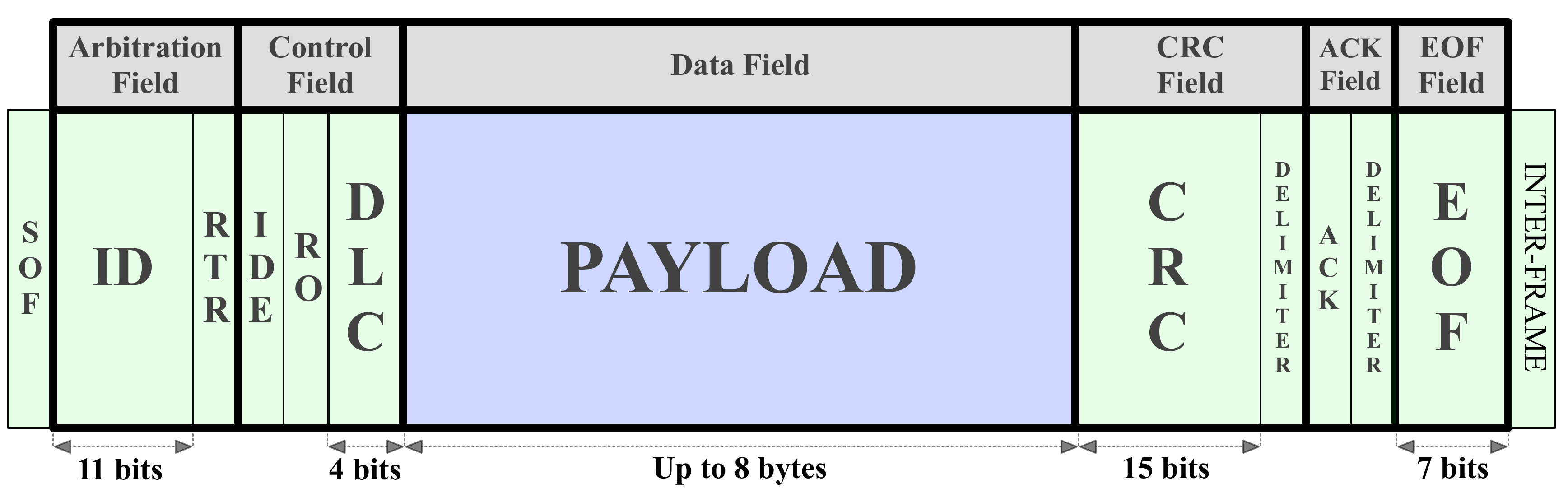}
	\caption{The various fields that make up a CAN frame.}
	\label{fig:canframe}	
\end{figure}

CAN is a multi-master serial bus with a typical speed of 500 kbit/s on which many of the operational and safety-critical in-vehicle functions are implemented. As shown in Figure~\ref{fig:canframe}, each CAN \emph{frame} has a unique ID that indicates its priority, determines its source and destination(s), and enables the receivers to decode its data content accordingly. If two or more CAN frames are to be transmitted simultaneously on a single bus that is in idle state, the frame with the highest priority (the lowest ID) is transmitted first. During this time, all other ECUs refrain from transmission, switch to \texttt{receiving} mode, and wait until the bus returns to the idle state. 

A single CAN frame can carry up to 8 data bytes of content, typically consisting of one or multiple signals that control different functions such as speed, steering-wheel angle, autonomous braking, indicators, and lamps. The permissible data content and transmission frequency of each CAN frame are specified by the vehicle manufacturer in the early phases of ECU development. For instance, a CAN frame with ID=0x12 and periodicity of 10 ms may belong to the engine-control ECU, in which case, it would carry data related to the engine speed and torque. A receiving ECU that is waiting for a frame with ID=0x12 examines the ID using a filtering mechanism and drops unwanted frames. Due to the strict specification of the ECU communication, the degree of randomness in IVN traffic is significantly lower than that of Internet traffic and classical computer networks.

CAN messages are broadcast and do not carry information about the sender or receiver, and thus the CAN protocol provides no mechanisms to verify their authenticity and integrity. Unlike MACs, which guarantee that the content of a message has not been maliciously altered in transit, the CRC field in a CAN frame only enables a receiver to detect bit flips caused by transmission errors.  

%CAN communication is highly periodic and modeled in a proprietary database called \textit{signal database} owned by the car manufacturer.
%\mynote{[NN: Explain a little more about how CAN bus works. Specifically, mention that each node listens to the bus for receiving certain message IDs and ignores messages with any other ID. The second point to mentions is that when two ECUs start transmitting with the same message ID, they both perform the arbitration phase and continue to send the rest of the message while none of them notices that the other node is transmitting at the same time! However, if after arbitration phase any of the two transmitting nodes send a bit $'1'$ while the other node is sending a bit $'0'$, since a bit $'0'$ is dominant, the transmitter of $'1'$ will send an error frame (because it has sent $'1'$ but read $'0'$ from the bus) and everyone stops transmitting including the other node. Make sure that you write this part in a formal CAN bus language]}

To gain access to the CAN traffic in a vehicle, the mandated On-Board Diagnostics (OBD-II) physical interface, typically situated under the steering wheel, is commonly used. OBD-II is a standardized diagnostic connector that gives access to various vehicle subsystems, and can be used as an entry point to monitor or inject CAN traffic~\cite{miller2013adventures}. 
%A well-defined set of Service IDs (SID), along with the parameters associated with the services, enable a testing equipment to communicate with in-vehicle ECUs. 
%Furthermore, it is a common practice to design and implement a system-check function for each individual ECU to check its current overall state or the state of any of its components. 
%Moreover, a diagnostic testing device may require running tests in order to find faulty ECUs. The UDS protocol provides a remote routine activation service for such tasks. 
Some ECUs have a set of \emph{service routines} that may be triggered during a diagnostic session by a testing device in order to identify errors. For instance, it is possible to establish a diagnostic session to the brake ECU and order it to 
fully engage the brakes by sending a request with the associated routine ID and parameters. This communication between the testing device and the ECUs is enabled by the Unified Diagnostic Services~(UDS) protocol, which is a standardized diagnostic communication protocol, commonly used in ECU communication. A wide range of services are available through UDS, such as ECU reset, read/write by identifier, diagnostic session control, and remote routine activation. Safety-critical service routines are typically protected and executable only under certain conditions. For instance, an engine ECU may perform a routine to \quotes{kill} a particular fuel injector only if the vehicle's speed is below some threshold. In order to unlock such services for authorized users, a request-response protocol, commonly referred to as \emph{Security Access}, must be initiated by the user device. This protocol ensures that only basic diagnostic commands, such as \texttt{ECU reset} and \texttt{listen-only mode}, are permissible for unauthorized users.

%However, as stated earlier, safety and security-critical service routines may be protected by the Security Access protocol and executable only under certain conditions. For instance, an engine ECU may perform a routine to \quotes{kill} a particular fuel injector only if the vehicle's speed is below a certain threshold.

\subsection{Related Work}
\label{subsec:relatedwork}
%\mynote{[NN: Sort chronologically, and categorize. Currently none is done!]}
Much of the related literature on in-vehicle attack detection lays particular emphasis on the well-defined specifications of CAN communication with respect to message periodicity and data content.~\citet{larson2008approach} propose a specification-based method to detect malicious communication that does not comply with the configuration parameters of the ECU and the CAN protocol specifications. After identifying eight potentially exploitable aspects of CAN communication, such as message frequency, data consistency, and data plausibility,~\citet{muter2010structured} propose a sensor-based detection method to detect abnormal events related to each aspect. 

An entropy-based approach is proposed in~\cite{muter2011entropy}, where the normal entropy of CAN traffic is initially modelled, and the entropy of subsequent traffic is then monitored such that a higher value indicates a higher level of coincidence in the communication. Since IVN traffic has strict specifications, manipulation in the frequency or payload of CAN frames is expected to increase the entropy, and may thus be detected by the proposed technique. 

Based on the assumption that each frame ID can be associated with only one transmitter on a single bus,~\citet{matsumoto2012method} suggest the use of the frame ID to both detect and prevent unauthorized message transmissions. According to the authors, an ECU should verify that its own messages are not present on the bus unless the ECU is in \texttt{sending mode}. If this is not the case, then it is likely that an unauthorized node is transmitting the message on behalf of the actual sender, in which case, the receiving ECU overwrites the message before the transmission is over. 

To detect message-injection attacks,~\citet{moore2017modeling} propose to model the inter-arrival time (or wait time) of periodic messages. Given that in a message-injection attack the adversary must send messages with a valid ID at a rate at least equal to the original message frequency, the proposed technique detects an attack whenever the inter-arrival time between two instances of a given message deviates from the expectation. 

Distinctively,~\citet{murvay2014source} exploit the physical characteristics of CAN frames (e.g., voltage) to fingerprint ECUs so as to achieve source detection. The authors argue that signals generated by CAN transceivers exhibit different patterns, even if the transceivers originate from the same manufacturer, due to peculiar differences in their physical properties. A follow-up study by~\citet{choi2016identifying} improves the fingerprinting capabilities by analyzing more CAN traffic features. In a similar fashion,~\citet{cho2016fingerprinting} propose to fingerprint CAN transceivers based on ECU clock behavior. Due to the lack of clock synchronization in CAN communication, the authors hypothesized that the \texttt{clock offset} and the \texttt{clock skew} of each CAN transceiver are inimitable because they depend only on each ECU's local crystal clock. Unlike other state-of-the-art methods, the clock-based technique has been shown capable of detecting more acute attacks. 

Even though the state-of-the-art IVN intrusion detection systems are capable of detecting many types of attacks, they suffer from certain drawbacks. In particular, most proposed methods require prior knowledge about the underlying IVN and ECU configurations, which may vary even in vehicles of the same model and year. Although such variety in configurations enables automakers to fulfill different geographical market needs, it makes it quite challenging to update the IDS models if existing approaches are to be deployed in vehicles. Furthermore, with regards to coverage of different attacks, the proposed techniques have not been shown capable of detecting advanced stealthy attacks. We will discuss these detection limitations in more detail in Section~\ref{sec:attackTaxonomy}. 
\section{Attacks and Vulnerabilities}
\label{sec:attackTaxonomy}
In this section, we introduce our new stealthy attack and present the various attack scenarios that we considered in our experiments after describing our adversary model.
%In this section, we discuss various attack surfaces in IVNs, introduce our adversary model, and present a taxonomy of the different attack scenarios that we consider in our experiments. Our aim is not to provide a comprehensive attack taxonomy, but rather to summarize and improve those offered in related work in this area~\cite{kleberger2011framework, miller2015remote, checkoway2011comprehensive, studnia2013survey, cho2016fingerprinting, Cho2017viden}.

The most common way of accessing a vehicle's internal network to inject attack messages is through its OBD-II port (see Section~\ref{subsec:can}). It is also possible to remotely access and attack the vehicle through other media such as the Internet, cellular networks, and Bluetooth systems~\cite{checkoway2011comprehensive, miller2015remote}. The in-vehicle infotainment system commonly has a CAN transceiver and provides external networking connectivity and is, as such, an attractive target for adversaries to gain access to the in-vehicle CAN bus.

\subsection{Adversary Model}
In our experiments, we assume that the adversary has either local or remote access to the IVN. Furthermore, we consider the worst possible situation in which a nefarious adversary aims to influence the vehicle behavior, potentially leading to immobilization, dangerous maneuvers, or even collision. 

Irrespective of the attack surface used to mount the attack, after having identified the CAN ID of the intended safety-critical message, the adversary's goal is to perform one of the following malicious tasks or a combination thereof: \textit{(i)} compromise the original sender ECU (hereinafter referred to as \emph{target} ECU) and maliciously manipulate the message payload before it is transmitted on the bus; \textit{(ii)} impersonate the target ECU by injecting the carefully crafted message in an indistinguishable manner; or \textit{(iii)} prevent the target ECU from sending the message by placing it in a \texttt{listen-only mode} via diagnostic commands. 

The cost and complexity of performing these tasks vary depending on the target vehicle's hardware, software, and attack surfaces.~\citet{foster2015fast} experimentally show how these factors, especially the ECU firmware, affect the ability of an adversary to inject malicious messages into the bus. In general, ECUs can be either \emph{partially} or \emph{fully} compromised.
%aims to either forge and inject a message with the same ID, or cease the original sender ECU from transmitting the message. in order to perform one of the following malicious tasks or a combination thereof: \textit{(i)} Maliciously manipulate the compromised ECU message payloads in attempt to impose her nefarious intents to the receiving ECUs; \textit{ii} If failed Impersonate an existing ECU by injecting carefully crafted messages in an indistinguishable manner; or \textit{(iii)} prevent a target ECU from sending certain messages by placing it in a \texttt{listen-only} mode via diagnostic commands. The cost and complexity of performing these tasks vary depending on the target vehicle's hardware, software, and attack surfaces.~\citet{foster2015fast} experimentally show how these factors, especially the ECU firmware, affect the ability of an adversary to inject malicious messages into the bus.

%As commonly pointed out in the literature~\cite{foster2015fast, miller2015remote, miller2013adventures, ofir2014remote, koscher2010experimental, checkoway2011comprehensive}, two types of adversaries are typically considered in the IVN domain: \emph{partially capable} and \emph{fully capable} adversaries.\\

\textit{\textbf{Partially compromised ECUs}}. If an ECU is \emph{partially} (or weakly) compromised, the adversary lacks access to its memory and can only listen to the communication on the bus. A partially compromised ECU may not be used to inject arbitrary messages, however, the adversary can forcibly prevent it from transmitting its normal messages by either putting it offline, or by putting the transceiver into \texttt{listen-only mode} via basic diagnostic commands.
ECU developers usually implement a set of service routines that may be triggered by a testing device in a workshop through a diagnostic session after the routine requirements have been met. As pointed out in Section~\ref{subsec:can}, safety-critical routines are well protected and execute only under certain conditions, such as the vehicle being at low speed or completely turned off. Importantly, a weakly compromised ECU may not be used to trigger or exploit such routines. An adversary may essentially use a weakly compromised ECU to sniff traffic or shut it down entirely to cause one or several other ECUs that rely on its messages to malfunction.

%Importantly, a weak adversary does not possess the required skills to reverse-engineer the ECU firmware and attempt to reprogram it. Hence, the biggest threat posed by a weak adversary is to sniff traffic on the bus or entirely shut down the target ECU to cause one or several other ECUs that rely on its messages to malfunction. \\
%It should be noted that the privilege of injecting is incapable of reprogramming the ECU in order to manipulate its message contents.

\textit{\textbf{Fully compromised ECUs}}. By contrast, if an ECU is \emph{fully} (or strongly) compromised, the adversary is assumed to have full memory access and the ability to inject crafted messages into the bus. An adversary who can fully compromise an ECU typically has comprehensive knowledge of its software and hardware specifications, and may be capable of reprogramming its firmware to add or remove ECU features. Moreover, with a fully compromised ECU, the adversary may obtain the Security Access seed/key generation algorithm by reverse-engineering the firmware to unlock the ECU and exploit its safety-critical routines. In fact, as demonstrated in recent automotive attacks~\cite{keen2017tesla, keen2016tesla, miller2015remote, checkoway2011comprehensive, koscher2010experimental}, it is not uncommon for an adversary to unlock a strongly compromised ECU using advanced diagnostic commands and bypass (or disable) existing software protection mechanisms.
\vspace{1em}
%Security Access and be able to unlock the ECU and use advanced diagnostic commands. This effectively implies that a fully capable adversary can potentially enable or disable ECU features including any existing security protection features. 

\begin{figure*}[!t]
    \centering
    \begin{subfigure}[t]{\textwidth}
        \centering
        \includegraphics[width=\linewidth]{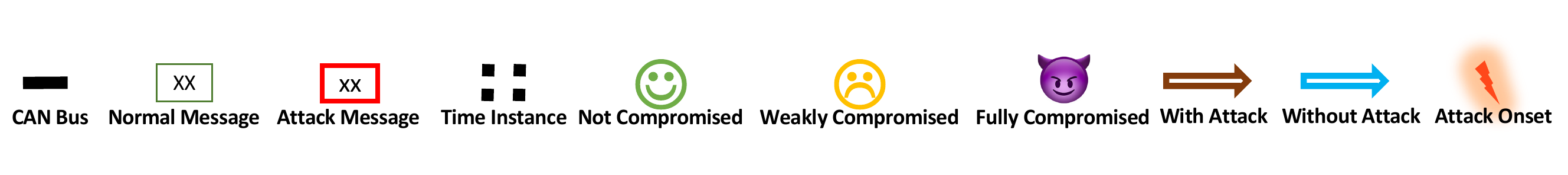}
        \label{fig:scheme-legends}
    \end{subfigure}%
    \vfill
    \begin{subfigure}[t]{\columnwidth}
        \centering
        \includegraphics[width=\linewidth]{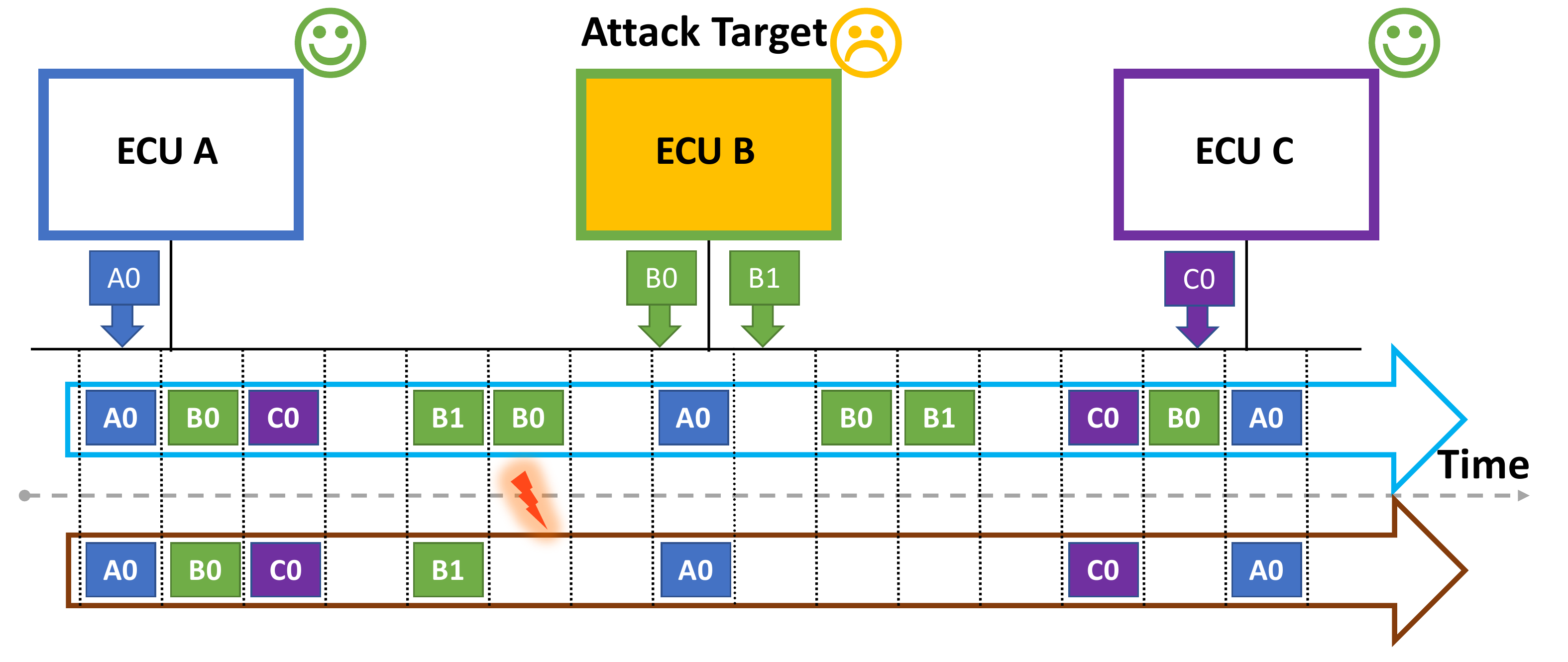}
        \caption{Suspension attack scheme}
        \label{fig:scheme-suspension}
    \end{subfigure}%
    \hfill
    \begin{subfigure}[t]{\columnwidth}
        \centering
        \includegraphics[width=\linewidth]{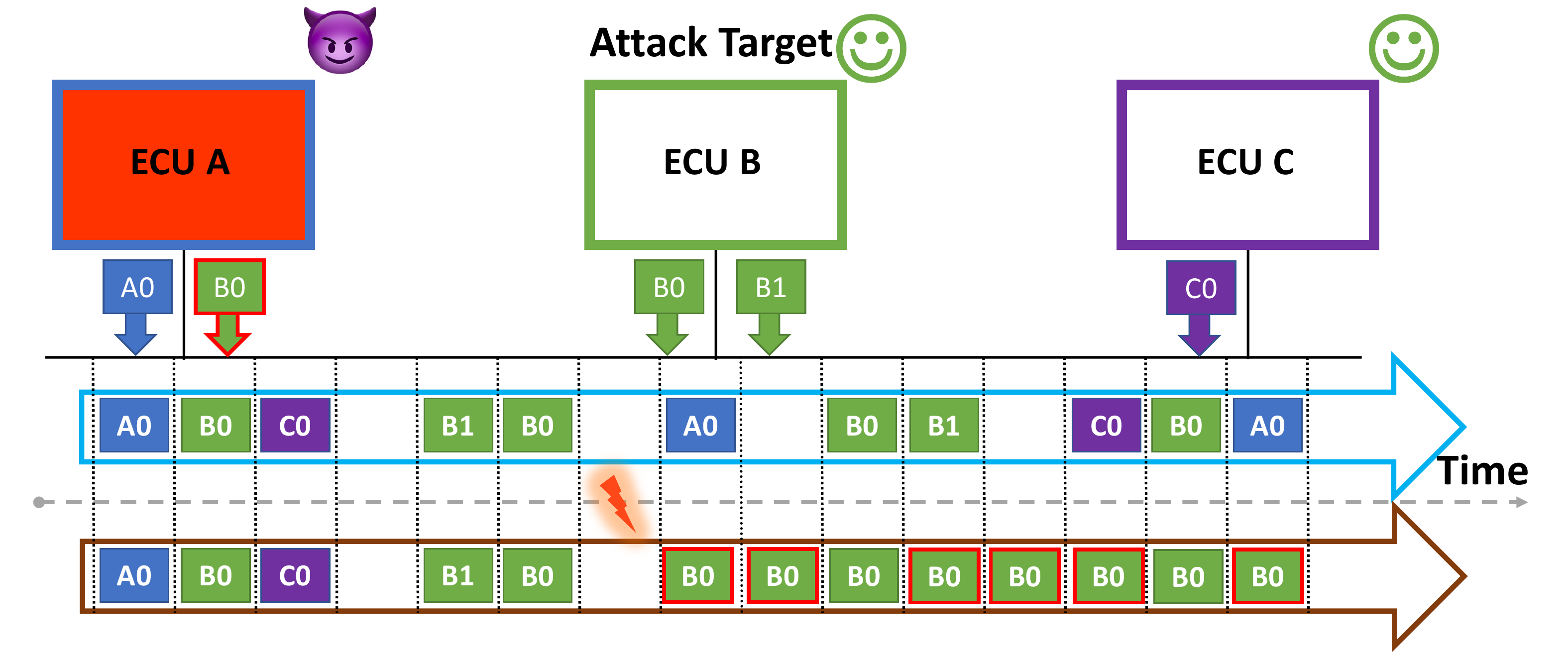}
        \caption{Fabrication attack scheme}
        \label{fig:scheme-fabrication}
    \end{subfigure}%
    \vfill
    \begin{subfigure}[t]{\columnwidth}
        \centering
        \includegraphics[width=\linewidth]{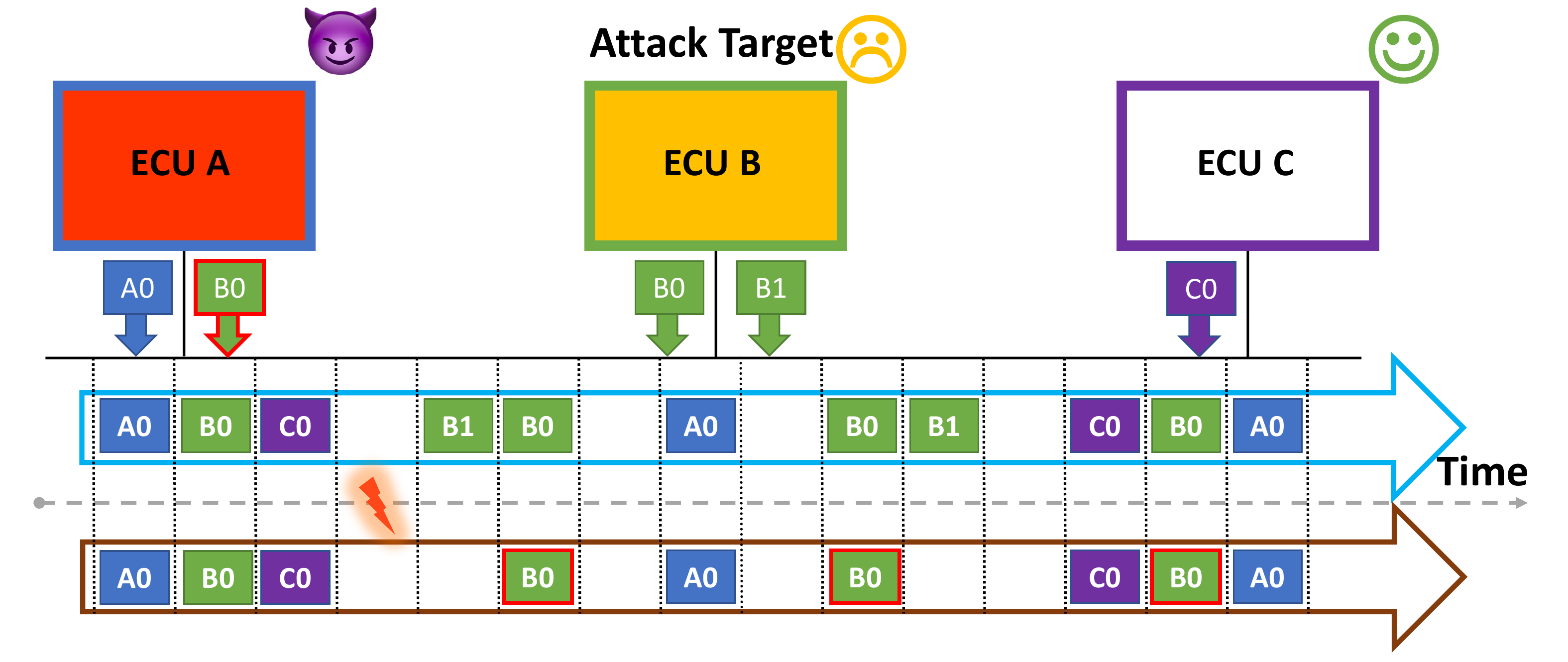}
        \caption{Masquerade attack scheme}
        \label{fig:scheme-masquerade}
    \end{subfigure}%
    \hfill
    \begin{subfigure}[t]{\columnwidth}
        \centering
        \includegraphics[width=\linewidth]{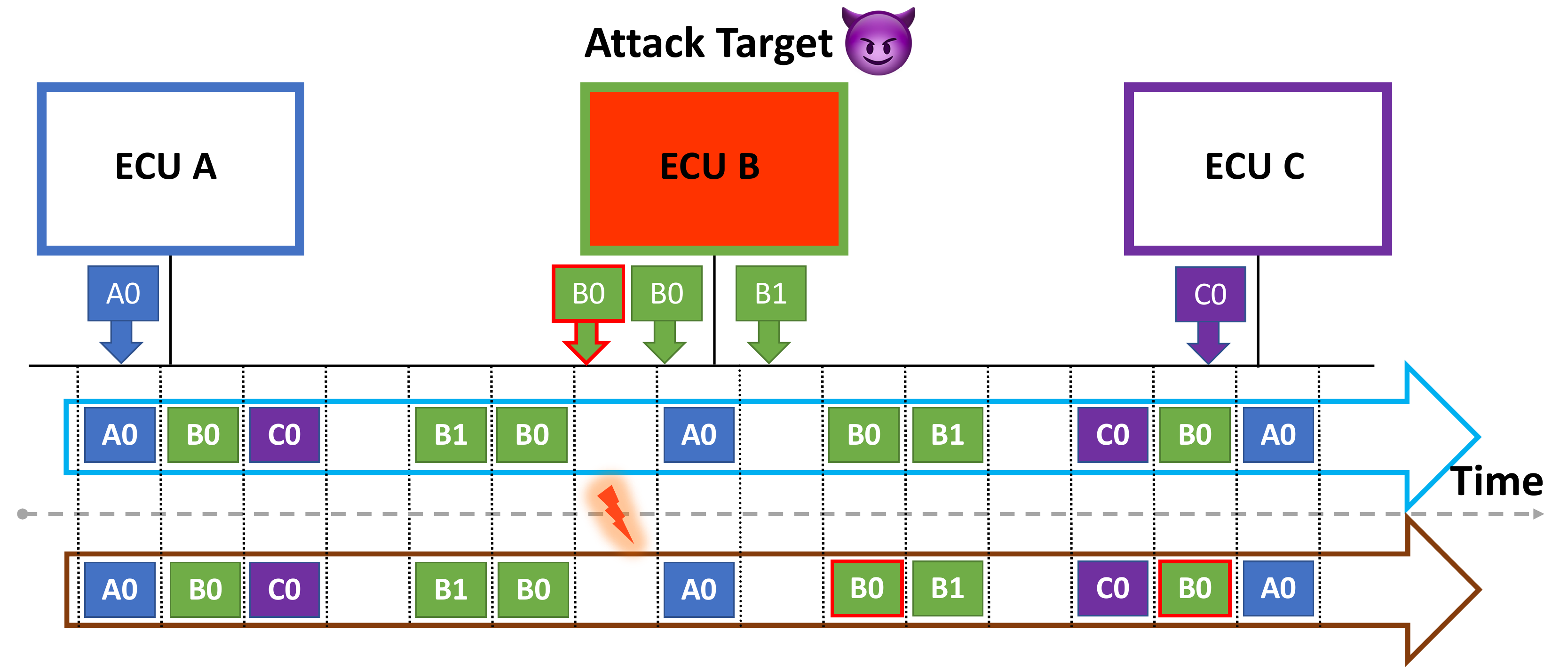}
        \caption{Conquest attack scheme}
        \label{fig:scheme-conquest}
    \end{subfigure}%
    \caption{Schematics for \emph{suspension} attack (\ref{fig:scheme-suspension}), \emph{fabrication} attack (\ref{fig:scheme-fabrication}), \emph{masquerade} attack (\ref{fig:scheme-masquerade}), and \emph{conquest} attack (\ref{fig:scheme-conquest}).}
    \label{fig:schematic-susp-fabric}
\end{figure*}

\subsection{Attack Scenarios}
\label{subsec:attackScenarios}
%There seems to be a consensus among researchers that there exist, effectively, three attack scenarios for in-vehicle networks:
In the literature, three attack scenarios for in-vehicle networks are documented: \emph{suspension}, \emph{fabrication}, and \emph{masquerade} attacks~\cite{Cho2017viden, cho2016fingerprinting, miller2015remote, miller2013adventures}, which are considered in our evaluation. In addition, as one of the contributions in this paper, we introduce a novel \emph{stealthy} attack which we refer to as the \emph{conquest attack}. 

In order to provide an intelligible explanation of the attacks, let us consider a simple CAN setup (Figure~\ref{fig:can-bus-prototype}) in which messages A0, B0/B1, and C0 are transmitted over different time intervals by ECUs $\mathbb{A}$, $\mathbb{B}$, and $\mathbb{C}$ respectively. 
%Message A0 is sent to receivers $\mathbb{B}$ and $\mathbb{C}$, message B0 is sent to $\mathbb{C}$, and message C0 is destined to both $\mathbb{A}$ and $\mathbb{B}$. 
For the sake of illustration, we consider B0 to be a sensitive message related to one or more safety-critical functions, and thus $\mathbb{B}$ is the target ECU in the attacks.

\textit{\textbf{Suspension attack.}} The suspension attack (Figure~\ref{fig:scheme-suspension}) is a type of Denial of Service (DoS) attack, where the weakly compromised ECU $\mathbb{B}$ ceases to send messages. In order to achieve this, the adversary manages, through a diagnostic session, to put the ECU into programming mode so that it is no longer able to transmit messages. As a result, the receivers that rely on incoming data from the suspended ECU may no longer function properly.
%Therefore, it can be achieved even on a weakly compromised ECU by simply putting it into establishing a programming session which makes it stop sending its normal messages. The impact can be quite devastating because the receivers may rely on incoming data from the suspended ECU and therefore may not function properly.\\
%\rawtext{The attacker can stop the ECU from transmitting certain messages by putting the ECU into sleep or receiving mode. The attacker is unable to inject spoofed messages.}

\textit{\textbf{Fabrication attack.}} In a fabrication attack, the adversary is incapable of compromising the target ECU, but is able to fully compromise another ECU ($\mathbb{A}$) on the bus and use it as a means to impersonate $\mathbb{B}$. Specifically, $\mathbb{A}$ is used to transmit forged messages with ID B0 at a higher frequency than $\mathbb{B}$ so that the receiving ECUs would receive conflicting B0 messages sent by both $\mathbb{A}$ and $\mathbb{B}$, except that the data received from $\mathbb{A}$ would be dominant due to higher frequency of transmission (see Fig~\ref{fig:scheme-fabrication}). Hence, the receiving ECUs would process the payload of the B0 messages overwhelmingly received from the adversary and react accordingly, potentially starting to malfunction. A typical scenario is the transmission of forged speedometer values at a higher pace than the original message, resulting in the speedometer behaving erratically on the dashboard and predominantly showing the speed attack values. 

For safety-related reasons, some automakers deploy the naive solution of associating with each message a counter value that is incremented by one on every new transmission, such that upon receiving two consecutive messages with non-matching counter values, the ECU initiates a fail-safe procedure or simply turns off. Unsurprisingly, it was demonstrated in~\cite{miller2013adventures} how it is possible to circumvent this type of protection and control the steering wheel of a Toyota Prius through a fabrication attack.
%\rawtext{The attacker is able to inject spoofed messages through a compromised ECU.}

\textit{\textbf{Masquerade attack.}} As can be noticed in the fabrication attack scenario, although the adversary impersonates the target ECU through a weakly compromised ECU, the target ECU will keep transmitting messages. This makes the fabrication attack ineffective whenever safety-related protection mechanisms on the receiving ECUs are in effect. These mechanisms are designed to identify and react to messages with contradicting signal values~\cite{miller2015remote}. 

In a masquerade attack (Figure~\ref{fig:scheme-masquerade}), on the other hand, the adversary has the additional power of weakly compromising the target ECU and suspending it from transmission, while immediately starting to transmit attack messages (B0) at the original frequency from a fully compromised ECU ($\mathbb{A}$) on the same bus. This way, the adversary can bypass possible protection mechanisms on the receiving ECUs, as the latter only process B0 messages coming from $\mathbb{A}$ and no violation in the arrival times of these messages occurs. 

As a case in point, in order to remotely attack and control a Jeep Cherokee,~\citet{miller2015remote} identified the message used by the Parking Assistant Module (PAM) ECU to control the steering wheel. However, unlike the previous vehicles they had successfully attacked (e.g., Toyota Prius), the fabrication of the message was ineffective on the Jeep. They observed that irregularities caused by sending fabricated messages cause the parking assistant system to get confused and go offline. However, they managed to get around the protection mechanism by stopping the PAM from sending its normal messages, thus switching from a fabrication attack to a masquerade attack. Specifically, they put the PAM into programming mode via a diagnostic session and then sent the real message from the fully compromised infotainment unit instructing the power steering ECU to turn the wheel.

\emph{The masquerade attack is not stealthy}. It is worth noting that since the target ECU is suspended by the adversary, it ceases to transmit \emph{all} of its messages (and typically ECUs transmit more than a single message), including B1; hence ECUs relying on receiving message B1 may start to malfunction. A masquerade attack may have impact on ECU~$\mathbb{A}$ as well, which may become overloaded in the likely case that it has fewer transmission buffers than the number of messages it has to transmit. This in turn would introduce problems such as significant priority inversion~\cite{davis2011controller} and non-abortable message transmission~\cite{di2012understanding} that consequently degrade the real-time performance of the CAN bus. In short, the masquerade attack is \emph{not} stealthy because it creates a mess in the CAN communication.

We deem an attack on IVNs \emph{stealthy} only if it causes no noticeable changes in CAN communication characteristics. Evidently, based on this definition of stealthiness, none of the attacks just described can be considered stealthy, since they all give rise to noticeable irregularities either in the timing behavior of IVN messages, or in the low-level physical properties of ECUs, or in both.

\textit{\textbf{Conquest attack.}} We introduce the \emph{conquest} attack, a truly stealthy type of attack, in which the adversary is able to evade both the security protection mechanisms and the state-of-the-art intrusion detection systems. As shown in Figure~\ref{fig:scheme-conquest}, in a conquest attack, the adversary directly \emph{conquers} the \emph{target} ECU by fully compromising it, which in none of the previous scenarios the adversary was able to achieve. In particular, the adversary is able to \emph{reprogram} the target ECU so that instead of having to compromise another node on the network to inject the intended malicious payload, the adversary directly manipulates the payload of the sensitive message B0, \emph{albeit only subtly}, thus forcing the corresponding safety-sensitive operation executed on the receiving ECU to function erroneously. Unlike all other scenarios we have described so far, this attack causes no changes in the normal behavior of any of the ECUs with respect to message frequency, clock offset, or clock skew behavior. Even at the payload level, a strategic adversary carrying out a conquest attack alters only a few bytes of data in a continuous stream of CAN frames. Such a stealthy attack may prove particularly effective against Advanced Driver-Assistance Systems (ADAS), such as forward collision warning, lane departure warning, and electronic stability control, that critically rely on the integrity of sensor values. 
%As a matter of fact, these systems are typically designed to initiate a fail-safe procedure whenever the sensor messages \mynote{in reality a majority of these systems use checksums (CRCs) to detect integrity related issues, rather than monitoring deviations in "timing" of messages. However, the probability of these systems having mechanisms for detecting deviations in "timing" behavior is not zero (e.g. start a degraded-mode if a message goes missing). Do we have to clarify this?}deviate from their original timing deadlines. 
Therefore, performing a conquest attack on such sensor values may have far-reaching consequences on the underlying IVN if the adversary manages to maliciously alter sensor data bytes in such a way that they fall into the normal range, yet deviate from the actual values received from the real environment.
A possible real-world conquest attack scenario is one where the adversary is able to reprogram the parking assistant module and directly send crafted steering wheel angles to the steering wheel ECU causing the wheel to behave erratically. In another hypothetical scenario, the adversary reprograms the engine control module and maliciously alters data bytes in the engine speed messages in such a way that the speed sensor readings remain in a reasonable range while not reflecting the actual speed of the engine. Consequently, functions that make use of these readings to make decisions based on the engine speed will be presented with misleading information.

We stress that a conquest attack is a serious one in no small part because it is feasible, stealthy, and potentially as damaging as the other three attacks. Furthermore, a conquest attack is not impractical; as a matter of fact, the currently available tools and techniques that were used in the most recent automotive attacks are sufficient for performing this type of attack. As a case in point,~\citet{miller2013adventures} were able to reverse-engineer the firmware of the parking assistant module on a Jeep Cherokee and obtain its Security Access algorithms and keys allowing them, in principle, to reprogram the ECU and eventually carry out a conquest attack. 

%However, since there were no IDS in the Jeep that can detect their attack and their goal was only to show the feasibility of remote attacks against vehicles, they settled for simply suspending the PAM module instead of reprogramming it for the sake of saving time and effort. In other words, had the Jeep been equipped with an in-vehicle intrusion detection system, they would have had to reprogram the PAM and perform a stealthy attack to avoid detection.

%\mynote{We may also add that he a conquest attack makes the compromised ECU to act like a faulty system rather an attacked one. This means that a smart adversary may reprogram the ECU to inject false information at certain times, or even stop showing faulty behavior in presence of a diagnostic testing device (e.g in workshop). The forensic value of information retrieved from an ECU that behaves faulty, may potentially be misleading and thus be negated.}

%Thus, making the attack detection difficult for the state-of-the-art IDS.  
%The objective of this attack is to manipulate the payload of a sensitive message (B0) sent by the ECU, thus preventing the related safety-critical function to operate properly. Unlike all other scenarios we have described so far, this attack causes no changes in the normal behavior of the ECU that launches the attack, including its message frequencies, clock offset and clock skew behavior. 

\subsection{Limitations of Existing Techniques}
\label{subsec:easeOfDetection}
Both suspension and fabrication attacks may cause severe impact on the vehicle, but since they cause drastic changes in the IVN traffic, they are trivial to detect by existing methods. In a masquerade attack, although no messages are sent at a higher frequency like in fabrication attacks, the weakly compromised target ECU is forced to refrain from transmission. The mere act of \emph{suspending} the target ECU leads to timing misbehaviors, which can be detected by network-level timing-based techniques. In the unlikely case where the adversary is capable of reproducing and retransmitting all messages originally sent by the target ECU, while respecting their complex scheduling structure, only fingerprinting techniques which recognize that messages are not sent by the target ECU can detect the attack. When performing a conquest attack, on the other hand, none of the IVN traffic characteristics is affected (see Figure~\ref{fig:scheme-conquest}). Therefore, the adversary may not only evade detection from the state-of-the-art techniques, but also possible safety-related protection mechanisms that monitor the incoming traffic for faulty behaviors. In our approach, described in the following section, we demonstrate that by monitoring the CAN data stream at the byte level, we are able to detect all types of attacks including the conquest attack (see~Table~\ref{tab:method-comparison}), 
without requiring to decode and understand the payload (i.e., the underlying signals). 
\begin{table}[!b]
\begin{tabular}{lcccc}
\hline
\multicolumn{1}{c}{\multirow{2}{*}{\textbf{Detection Approach}}} & \multicolumn{4}{c}{\textbf{Attack Type}}               \\ \cline{2-5} 
\multicolumn{1}{c}{}                                             & Susp. & Fab. & Masq. & Conq. \\ \hline
LLP\textbf{*}~\cite{murvay2014source,choi2016identifying,cho2016fingerprinting,Kneib2018}                &\cmarkg            &\cmarkg             &\cmarkg            &\xmarkr          \\ \hline
TBS\textbf{**}~\cite{larson2008approach, matsumoto2012method, muter2011entropy, moore2017modeling}                &\cmarkg            &\cmarkg             &\xmarkr            &\xmarkr           \\ \hline
\acro{}              &\cmarkg            &\cmarkg             &\cmarkg            &\cmarkg           \\ \hline
\end{tabular}
\caption{Detection capabilities: \acro{} vs. the state-of-the-art.}
\label{tab:method-comparison}
\raggedright\footnotesize{\ \textbf{*} Deviations in Low-Level Properties of CAN traffic (LLP).\\} 
\raggedright\footnotesize{\textbf{**}\ Deviations in Timing Behavior or Specifications of CAN traffic (TBS).}
\end{table}

\section{Methodology}
\label{sec:methodology}
\begin{figure}[!t]
\centering
\begin{subfigure}[t]{\columnwidth}
\includegraphics[width=\linewidth]{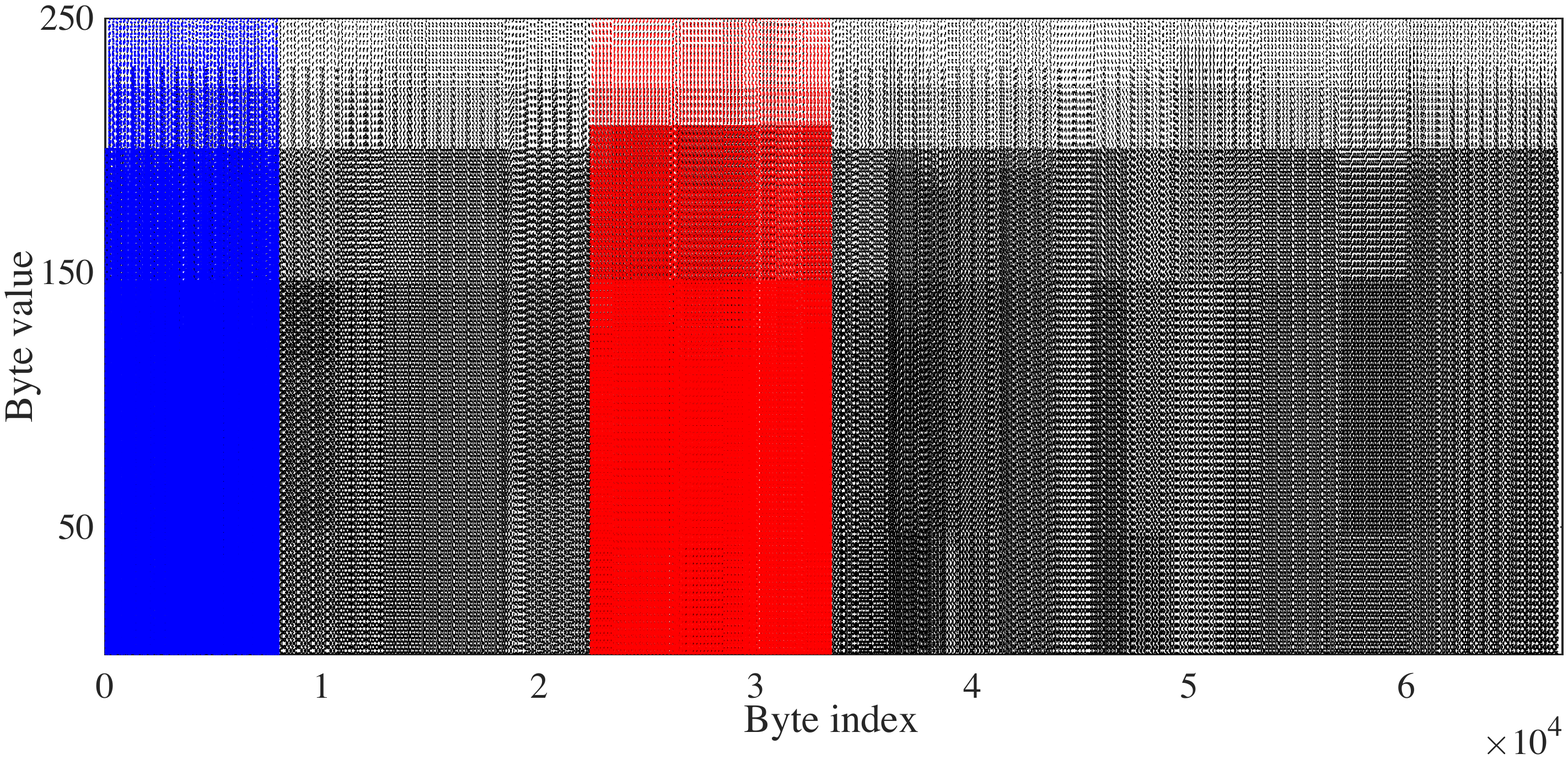}
\caption{CAN traffic}
\label{fig:can_traffic}
\end{subfigure}
%\hspace{1em}
\vfill
\begin{subfigure}[t]{\linewidth}
\centering
\includegraphics[width=\columnwidth]{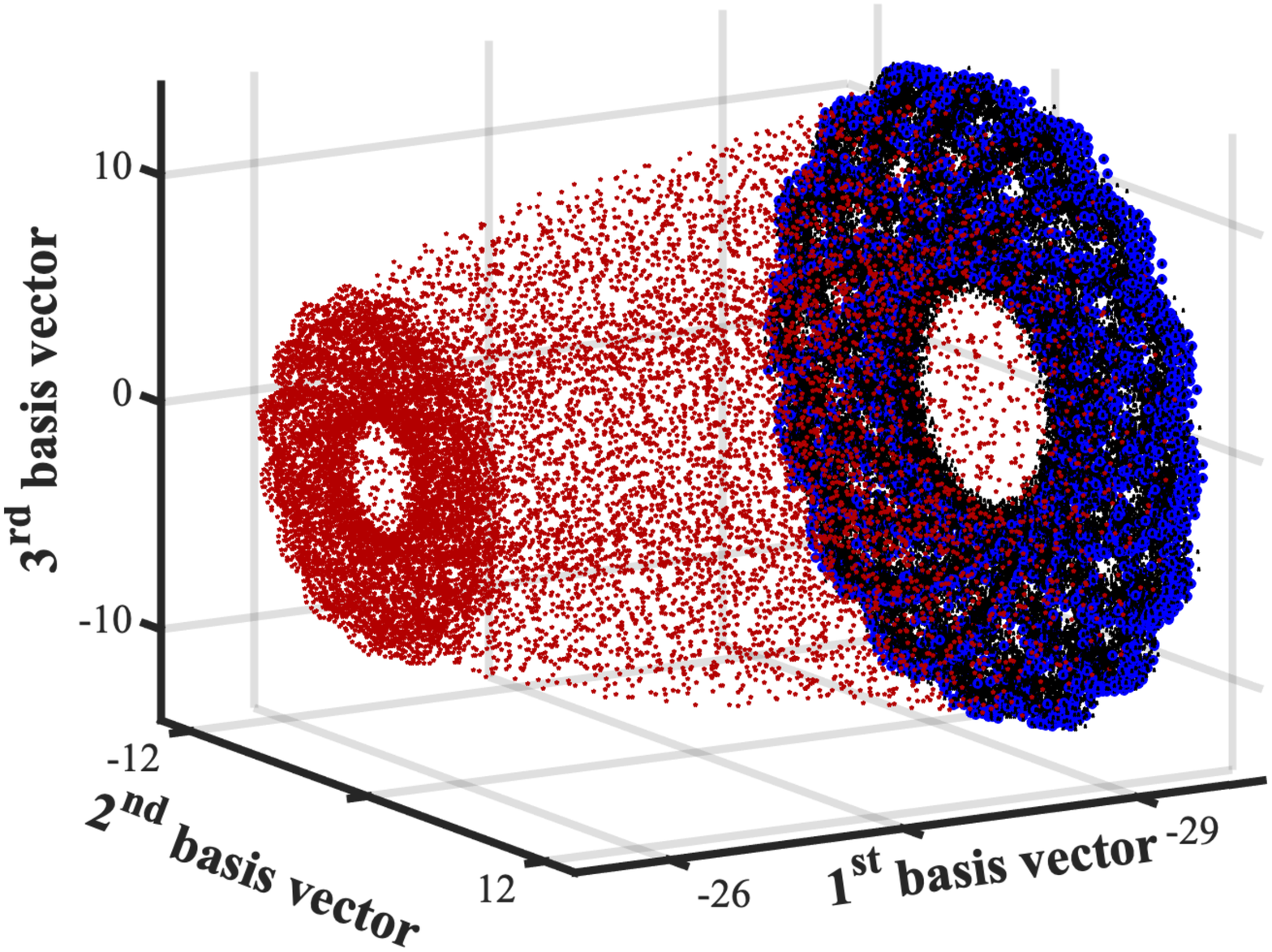}
\caption{Signal subspace.}
\label{fig:masq_3D}
\end{subfigure}
\hfill
\caption{Visualization of the departure of vehicular dynamics from normal behavior while the IVN is undergoing a masquerade attack.}
\end{figure}

In this section, we elucidate our attack-detection methodology. We present a brief overview of PASAD and then highlight the main adjustments we have introduced to develop \acro{} for in-vehicle networks. For a comprehensive treatment of PASAD, the reader may refer to~\cite{aoudi2018pasad}. 
In a nutshell, with \acro{}, we aim to detect all types of attacks on IVNs described in Section~\ref{subsec:attackScenarios} by monitoring CAN traffic at the \emph{payload level} to identify \emph{structural changes} attributable to malicious payload manipulation, where we process the data fields in the stream of frames transmitted over the CAN bus one byte at a time.
A key enabling property for \acro{} is the regularity of vehicular dynamics, which seem to follow a pattern with military precision. Given that this regularity is highly reflected in the CAN traffic (see Figure~\ref{fig:can_traffic}), \acro{} lends itself to specification-agnostic detection of attacks on IVNs. The main motivation for using \acro{} is its inherent capability of detecting \emph{slight} structural changes in the payloads by recognizing unusual byte sequences that were unseen during training, no matter whether the individual maliciously altered bytes fall into the normal range or not.

\subsection{PASAD: Process-Aware Detection}
PASAD is a model-free Process-Aware Stealthy-Attack Detection mechanism that has recently been proposed in the ICS domain to determine if the underlying process is drifting from historical normal behavior by continuously monitoring physical sensor measurements. The method is based on a time-series analysis technique known as \emph{singular spectrum analysis}, whereby signal information representing the deterministic behavior of the underlying dynamical system is extracted from a time series of sensor measurements. The method takes as input a time series $\mathcal{T}$ of sensor measurements and works in two phases: an offline learning phase and an online detection phase. In the learning phase, a subseries of the monitored signal used for training is embedded in a vector space, referred to as the \emph{trajectory space}. Then, a \emph{signal subspace} is identified through a mathematical procedure, onto which \emph{training vectors} are projected. Owing to the regularity in the underlying system behavior, the projected vectors form a \emph{cluster} in the signal subspace, thereby defining the normal behavior. Afterwards, during an online detection phase, at every iteration, a \emph{test vector} is composed by incorporating the most recent sensor value. A departure score is then iteratively computed by measuring the distance between the most recent test vector and the centroid of the determined cluster. Finally, an alarm is generated whenever the score crosses a prescribed threshold.

\noindent\textbf{\textit{Learning Phase.}}
Formally, in the learning phase, an initial subseries of $\mathcal{T}$ of length $N$ is unfolded into a \emph{trajectory} matrix $\B = \left [\vc{b}_\ms{1} : \vc{b}_\ms{2} : \cdots : \vc{b}_\ms{K}\right ]$ by forming $K$ $L$-lagged vectors $\vc{b}_i$, where $L$ is called the lag parameter, $1 \leq i \leq K,\text{ and } K = N-L+1$. Then, the \emph{singular value decomposition} of $\B$ is performed to obtain an orthonormal set of $L$ eigenvectors $\vc{u}_\ms{1}, \vc{u}_\ms{2}, \cdots, \vc{u}_\ms{L}$ of the covariance matrix $\B\B^\trans$. A matrix $\Um = \left [\vc{u}_\ms{1} : \vc{u}_\ms{2} : \cdots : \vc{u}_\ms{r}\right ]$ is then formed, whose columns are the $r<L$ leading eigenvectors, where $r$ is the so-called \emph{statistical dimension}. The training vectors $\vc{b}_\ms{i}, 1 \leq i \leq K,$ are then projected onto the signal subspace spanned by the column vectors of $\Um$, and the centroid of the cluster they form is computed as $\vc{\tilde{c}}=\Um^\trans\vc{c}$, where $\vc{c}$ is the sample mean of the training vectors.

\noindent\textbf{\textit{Detection Phase.}} At every iteration during the detection phase, a \emph{departure score} is computed for the most recent lagged vector $\vc{b}_\ms{j}, j > K$. This is done by computing the squared Euclidean distance between the centroid $\vc{\tilde{c}}$ and the most recent test vector $\vc{b}_\ms{j}$ as 
\begin{equation}
\label{eq:distance}
D_j=||\vc{\tilde{c}}-\Um^\trans\vc{b}_j||^2. 
\end{equation}

\subsection{\stacro{}: CAN-Aware Detection}
PASAD is not applicable to IVNs out of the box. We have developed \acro{} for in-vehicle networks by making two important changes in PASAD: a CAN-traffic modeling mechanism and a modified procedure for computing the departure score.

\noindent\textbf{\textit{Time series of IVN Traffic.}} In \acro{}, the CAN traffic is modeled as a time series $\mathcal{T} = b_1,b_2,\cdots,b_\ms{N},b_\ms{N+1},\cdots$ of (the integer representation of) bytes extracted from the payloads of consecutive messages, such that if the payload of a message $m_\ms{j}$ on the CAN bus consists of bytes $b_\ms{i},b_\ms{i+1},b_\ms{i+2}$, then $b_\ms{i+3},b_\ms{i+4}$ would belong to the following message $m_\ms{j+1}$ whose payload contains only 2 bytes of data. The rationale behind processing individual bytes is that the CAN data field, which may contain up to 8 bytes of signal data, has a variable length, yet always contains a multiple of one byte. Subsequently, the vectors $\vc{b}_i = (b_\ms{i}, b_\ms{i+1}, \cdots, b_\ms{i+L-1})^\trans$ are composed to construct the trajectory matrix and perform the training.
\iffalse
\begin{equation}
\B =
			\begin{bmatrix}
			b_1 & b_2  & \dots & b_\ms{K} \\
			b_2 & b_3  & \dots & b_\ms{K+1} \\
			\vdots & \vdots & \ddots & \vdots \\
			b_\ms{L} & b_\ms{L+1}  & \dots & b_\ms{N} 
			\end{bmatrix}.
\end{equation}
\fi

\begin{figure}[t]
    \centering
    \begin{subfigure}[t]{.49\columnwidth}
        \centering
        \includegraphics[width=\linewidth,height=1.1in]{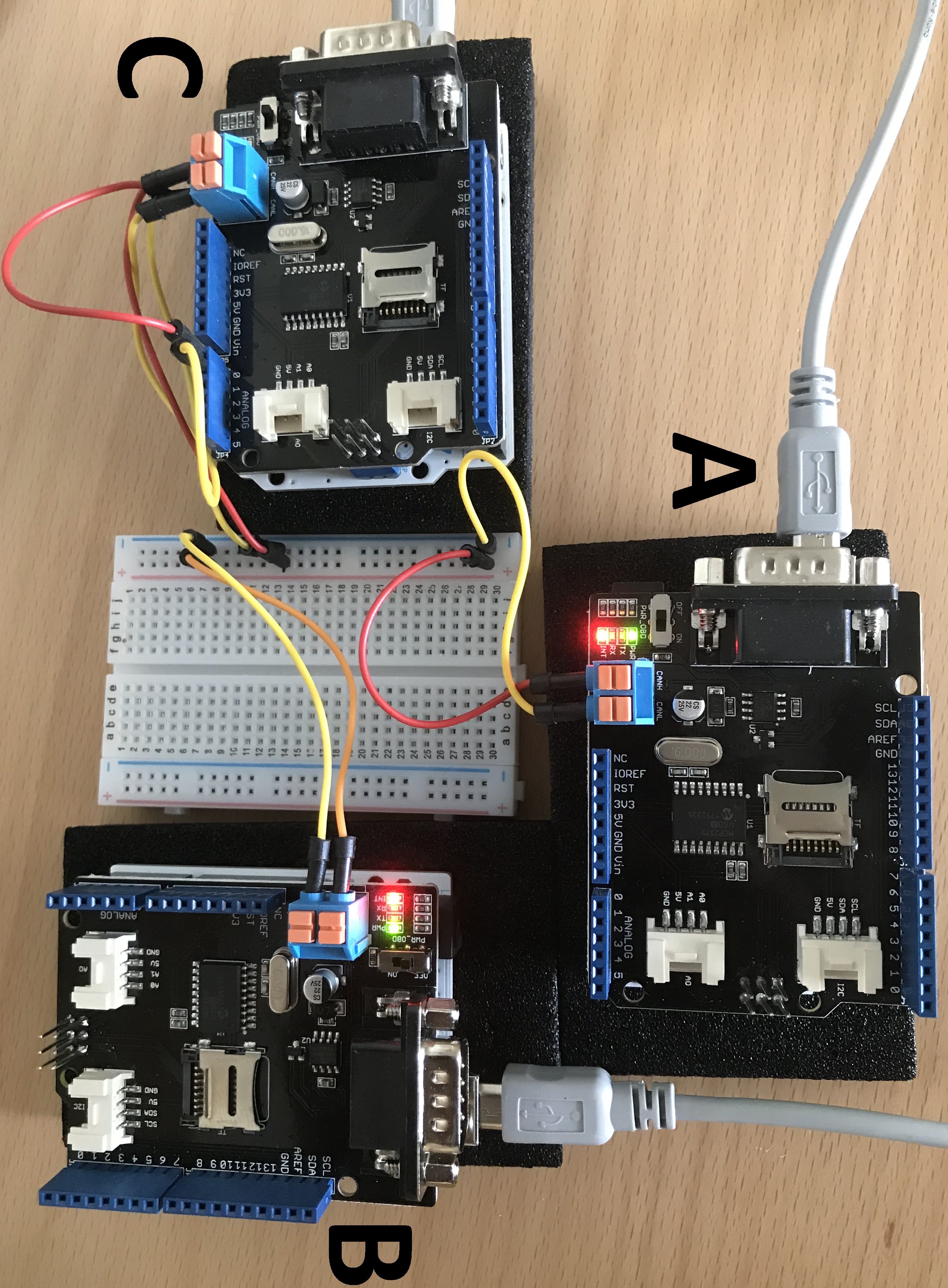}
        \caption{CAN bus prototype}
        \label{fig:can-bus-prototype}
    \end{subfigure}%
    \hfill
    \begin{subfigure}[t]{0.49\columnwidth}
        \centering
        \includegraphics[width=\linewidth,height=1.1in]{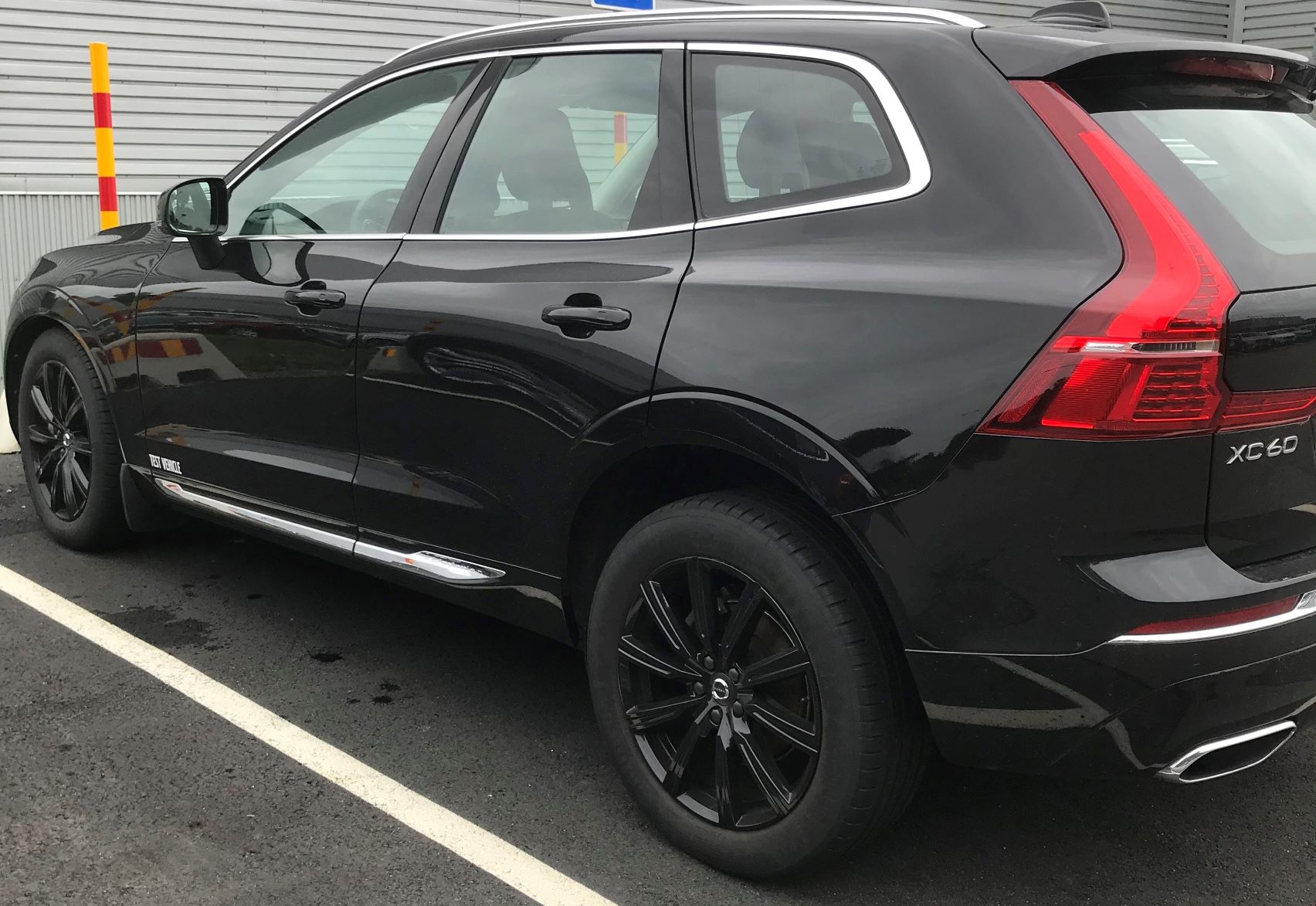}
        \caption{2018 Volvo XC60}
        \label{fig:xc60}
    \end{subfigure}%
   \vfill
    \begin{subfigure}[t]{\columnwidth}
        \centering
        \includegraphics[width=0.5\linewidth,height=1.1in]{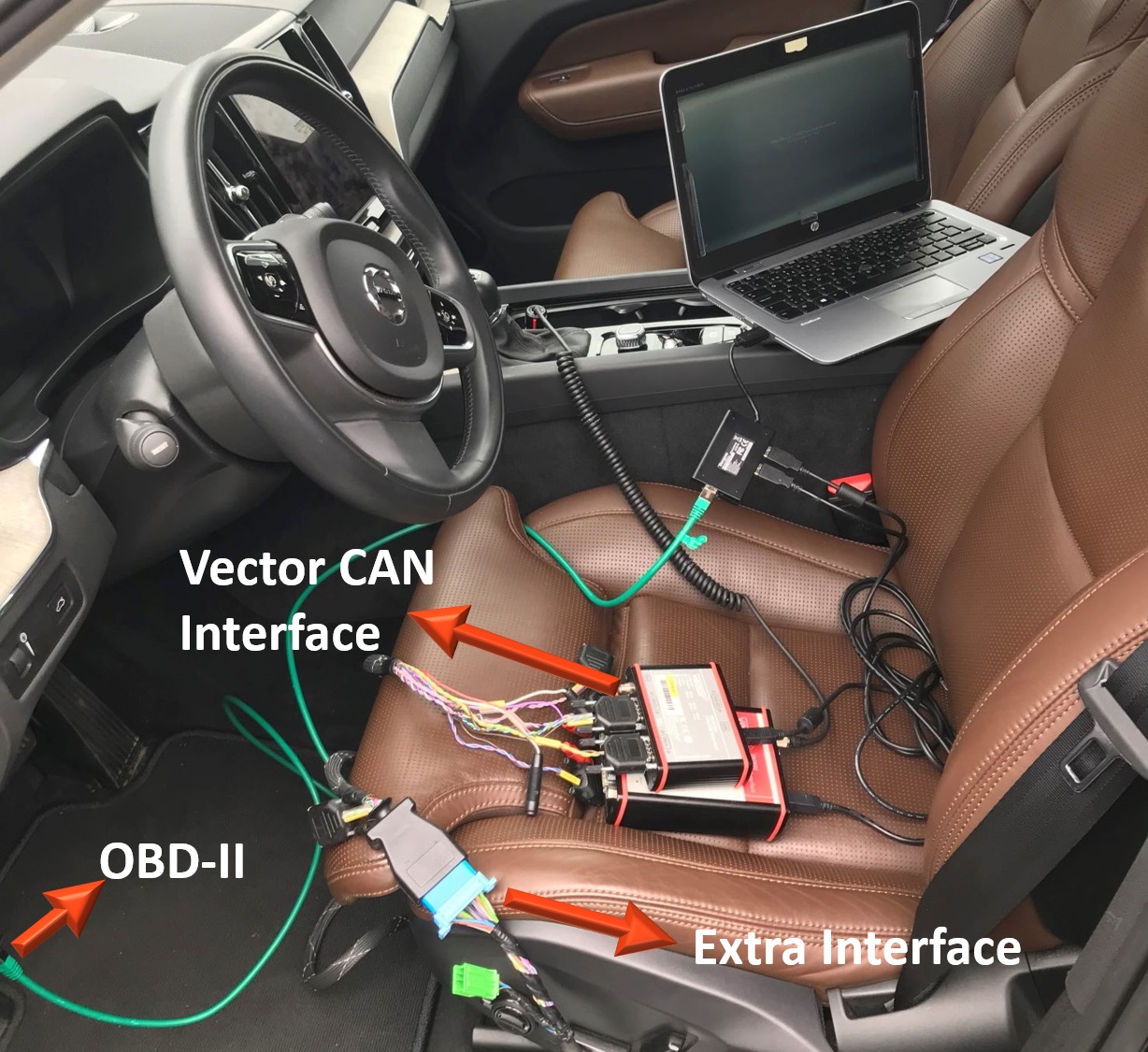}
        \caption{In-car OBD connection setup}
        \label{fig:in-car-obd}
    \end{subfigure}%
    \caption{Different setups used for evaluation.}
    \label{fig:evaluation-setups}
\end{figure}

\noindent\textbf{\textit{New Departure Score.}} Empirical evaluation on CAN traffic showed that during the detection phase the departure scores computed according to Eq.~(\ref{eq:distance}) were poorly scaled and did not measure up to the evident departures observed by visualization (see Figure~\ref{fig:masq_3D}). This result may largely be attributed to the fact that the monitored signal consists of integer values spanning a bounded range, unlike the typical application of PASAD on continuous real-valued sensor measurements. Therefore, in \acro{}, we determine the departure score for every test vector by computing the squared \emph{weighted} Euclidean distance from the centroid, where the weights are determined by the ratio of each eigenvalue to the total sum of eigenvalues associated with the $r$ eigenvectors determined in the learning phase. Specifically, we recompute the departure score in Eq.~(\ref{eq:distance}) as $\tilde{D}_j = ||\mathbf{W}(\vc{\tilde{c}}-\Um^\trans\vc{b}_j)||^2$
such that $\mathbf{W}$ is an $r$-dimensional diagonal matrix whose $i^\ms{th}$ diagonal entry is defined as $e_i/\ms{(\sum_i^r e_i)}$, where $e_i$ is the eigenvalue corresponding to the $i^\ms{th}$ eigenvector $\vc{u}_i$, for $1\leq i\leq r$. Finally, an alarm is generated whenever $\tilde{D}_j\geq \theta$, where $\theta$ is a prespecified threshold.
Figure~\ref{fig:can_traffic} shows a CAN traffic generated in a simulation setting using a testbed (see Section~\ref{subsec:evaluation_setups}). The initial subseries highlighted in blue corresponds to attack-free CAN traffic used in the learning phase to identify signal information that is representative of the normal behavior of the underlying vehicular communication. Once the learning phase is complete, a detection phase is initiated, in which a departure score is computed for every new traffic instance. The subseries of CAN traffic highlighted in red corresponds to malicious traffic injected by the attacker during a masquerade attack (described in Section~\ref{subsec:attackScenarios}).

\citet{aoudi2018pasad} showed that the signal subspace is \emph{isomorphic} to a low-dimensional Euclidean space, which enables the visualization of the projected vectors in $\mathbb{R}^\mathsmaller{3}$. Figure~\ref{fig:masq_3D} displays the vectors corresponding to the CAN traffic presented in Figure~\ref{fig:can_traffic} after they have been projected onto the identified signal subspace. Owing to the highly regular behavior of IVNs, as Figure~\ref{fig:masq_3D} depicts, the projected \textcolor{blue}{training vectors} occupy a firmly bounded region in the signal subspace and thereby form a cluster. Under normal conditions, the projected test vectors fall close to a cluster of training vectors. Under attack conditions, on the other hand, \textcolor{myred}{anomalous test vectors} are forced to lie further away from the cluster and the computed distance is presumed to increase, signaling that the IVN is departing from the normal behavior. This attack-indicating departure from normal behavior is expressed succinctly in Figure~\ref{fig:masq_3D}.~\citet{aoudi2018pasad} further introduced what they refer to as the \emph{isometry trick} to speed up the computations during the necessarily real-time detection phase. Hence, the detection procedure is quite fast, requiring a single matrix multiplication for every distance computation~\cite{aoudi2018pasadmidbro}.

\noindent\textit{\textbf{Limitations.}} Notwithstanding the importance of CAN communication, which carries most of the critical signals in modern vehicles, \acro{} only handles CAN traffic to detect attacks on IVNs, at least as far as tested. As \acro{} essentially performs spectral analysis of time series, the monitored messages also need to be periodic or frequently transmitted over the CAN bus, which is the case for most signals. The way \acro{} would handle non-periodic traffic depends on the type of messages being transmitted (i.e., somewhat common event-driven messages or very rare messages). As discussed in the attack-free experiment in Section~\ref{subsec:normalTrafficEvaluation}, while driving a real car, we intentionally emulated passenger-triggered controls and observed that such benign activity would be gracefully handled by \acro{}. In the case of rare events (e.g., emergency messages), \acro{} may require the complementary solution of blacklisting or whitelisting the messages for appropriate handling. Furthermore, as \acro{} monitors for changes in behavior, it may not react promptly to attacks in which the adversary manages to drive the vehicle to an unsafe state by making slow normal-looking changes at the payload level (e.g., linearly increasing engine speed) through a conquest attack.

\section{Evaluation}
\label{sec:experiment}
In order to evaluate the effectiveness of \acro{}, we conduct a series of experiments on a \textit{real} vehicle, a CAN bus prototype, and CAN traffic captured from two additional vehicle brands. In the first set of experiments, we demonstrate the capability of our approach to detect all attacks described in this paper, including the stealthy \emph{conquest} attack. In the second set of experiments, we examine the behavior of \acro{} when performing under attack-free conditions. We begin with a description of the evaluation setups and the data used in validating our approach.

\subsection{Evaluation Setup}
\label{subsec:evaluation_setups}
\textit{\textbf{CAN bus Prototype.}}
As shown in Figure~\ref{fig:can-bus-prototype}, we build an experimental CAN bus prototype consisting of three ECUs that communicate on a single bus. Each ECU consists of a SeeedStudio CAN-bus shield plugged on top of an Arduino Uno PCB. The CAN-bus shield adopts the Microchip MCP2515 SPI, a widely used CAN controller, and the Microchip MCP2551 CAN transceiver. The Arduino Uno is a microcontroller that is based on the ATmega328P with a 16 MHz quartz crystal clock, 32KB flash memory, and 2KB SRAM. ECU $\mathbb{A}$ was programmed to transmit message 0x1C every 30\textit{ms}, and ECU $\mathbb{B}$ was programmed to transmit messages 0x01 and 0x05 every 15\textit{ms} and 25\textit{ms} respectively, in addition to a third ECU ($\mathbb{C}$) which was programmed to capture the CAN traffic. To produce a realistic behavior, we simulate three messages with transmission frequencies and payloads (i.e., signals) that are similar to arbitrarily chosen safety-critical messages from a 2018 Volvo XC60. The CAN bus prototype was set up to operate at 500Kbps, which is the typical bit rate for in-vehicle high-speed CAN buses.

%To show the practicality of \acro{} in existing in-vehicle ECUs with low to medium demand of performance, we chose Arduino Uno,

\begin{figure}[!b]
    \centering
    \begin{subfigure}[t]{0.49\columnwidth}
        \centering
        \includegraphics[width=\linewidth]{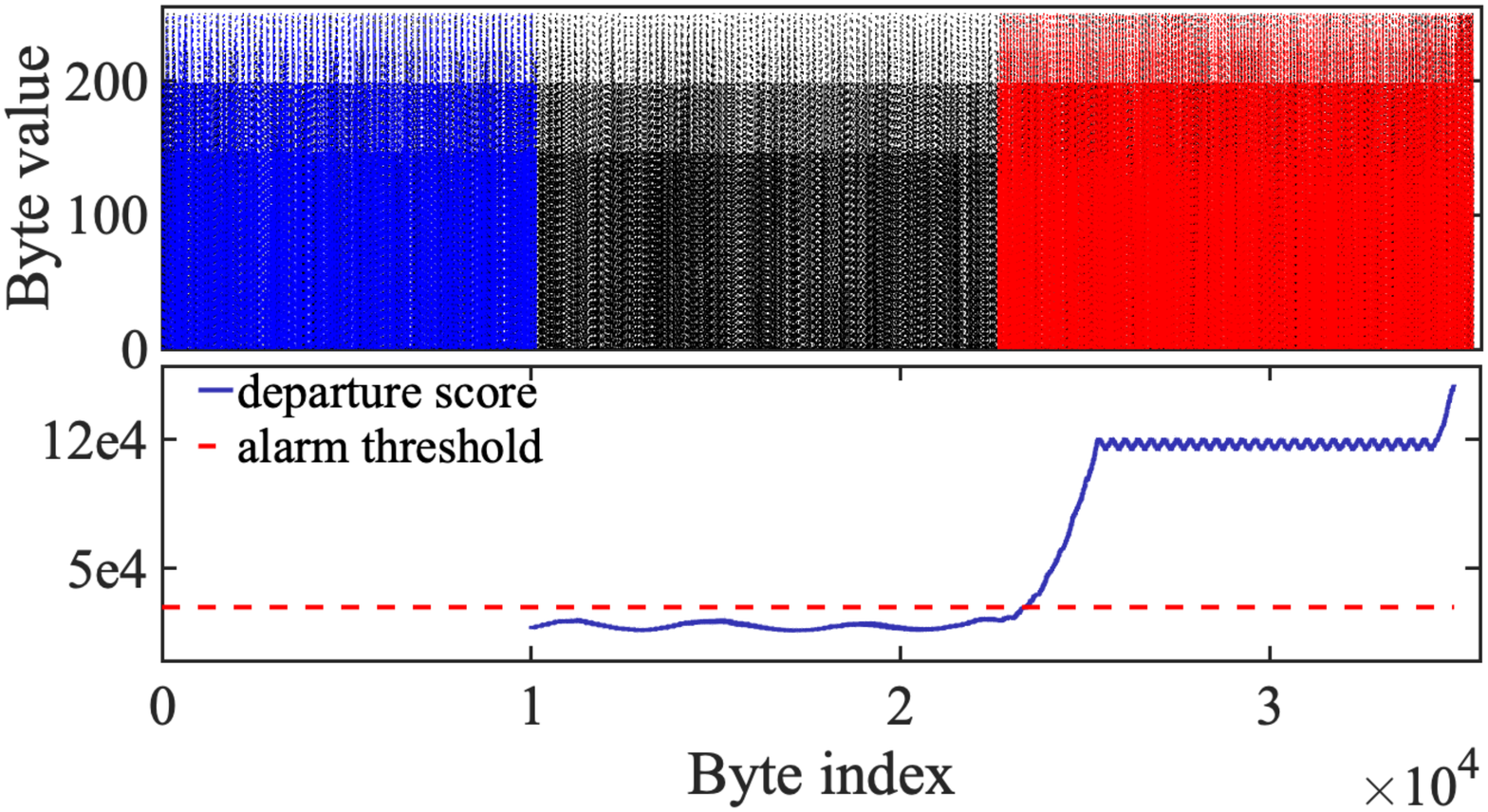}
        \caption{Suspension on prototype}
        \label{fig:suspension_attack_prototype}
    \end{subfigure}
    \hfill
    \begin{subfigure}[t]{0.49\columnwidth}
        \centering
        \includegraphics[width=\linewidth]{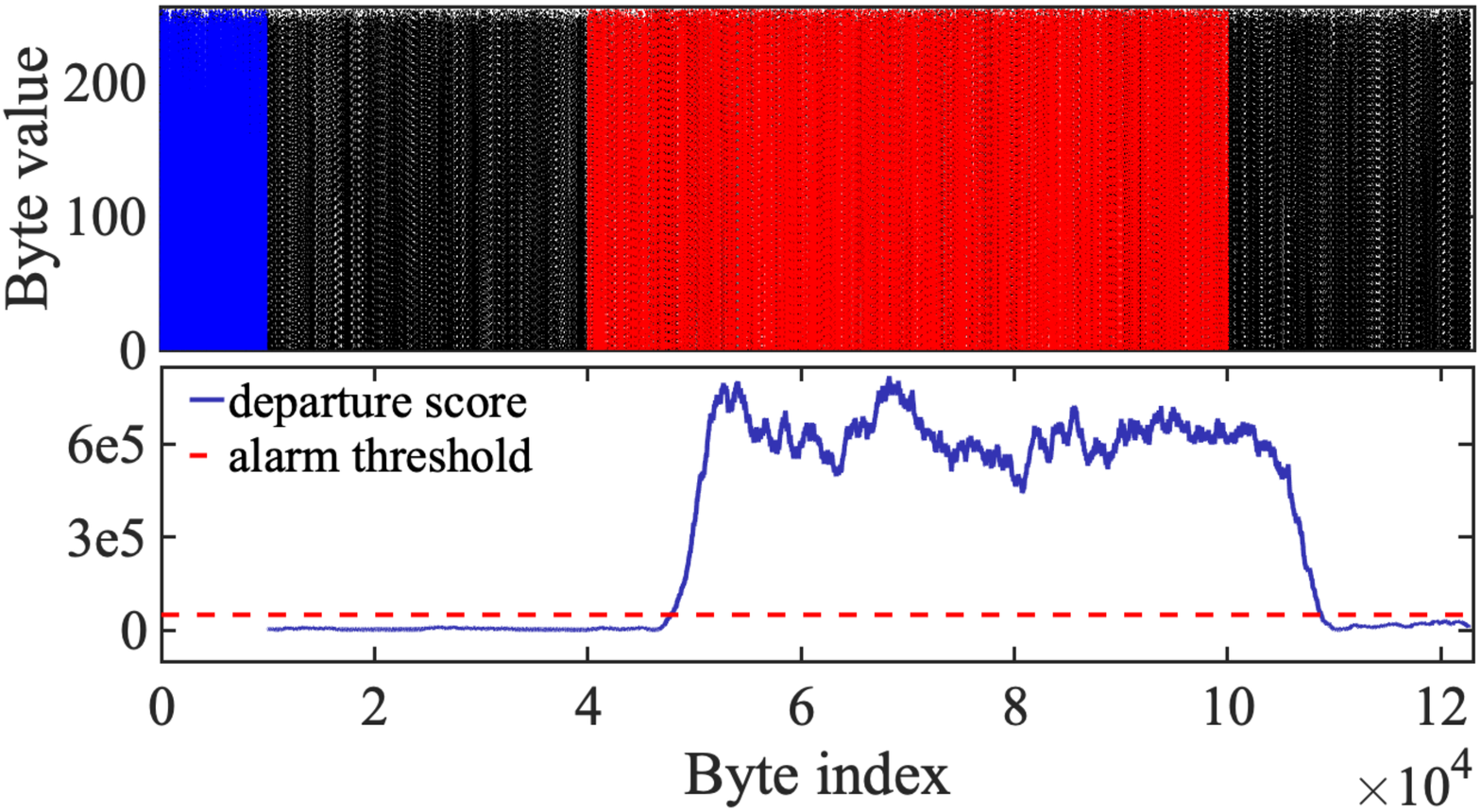}
        \caption{Suspension on Volvo XC60}
        \label{fig:suspension_attack_xc60}
    \end{subfigure}
    \vfill
    \begin{subfigure}[t]{0.49\columnwidth}
        \centering
        \includegraphics[width=\linewidth]{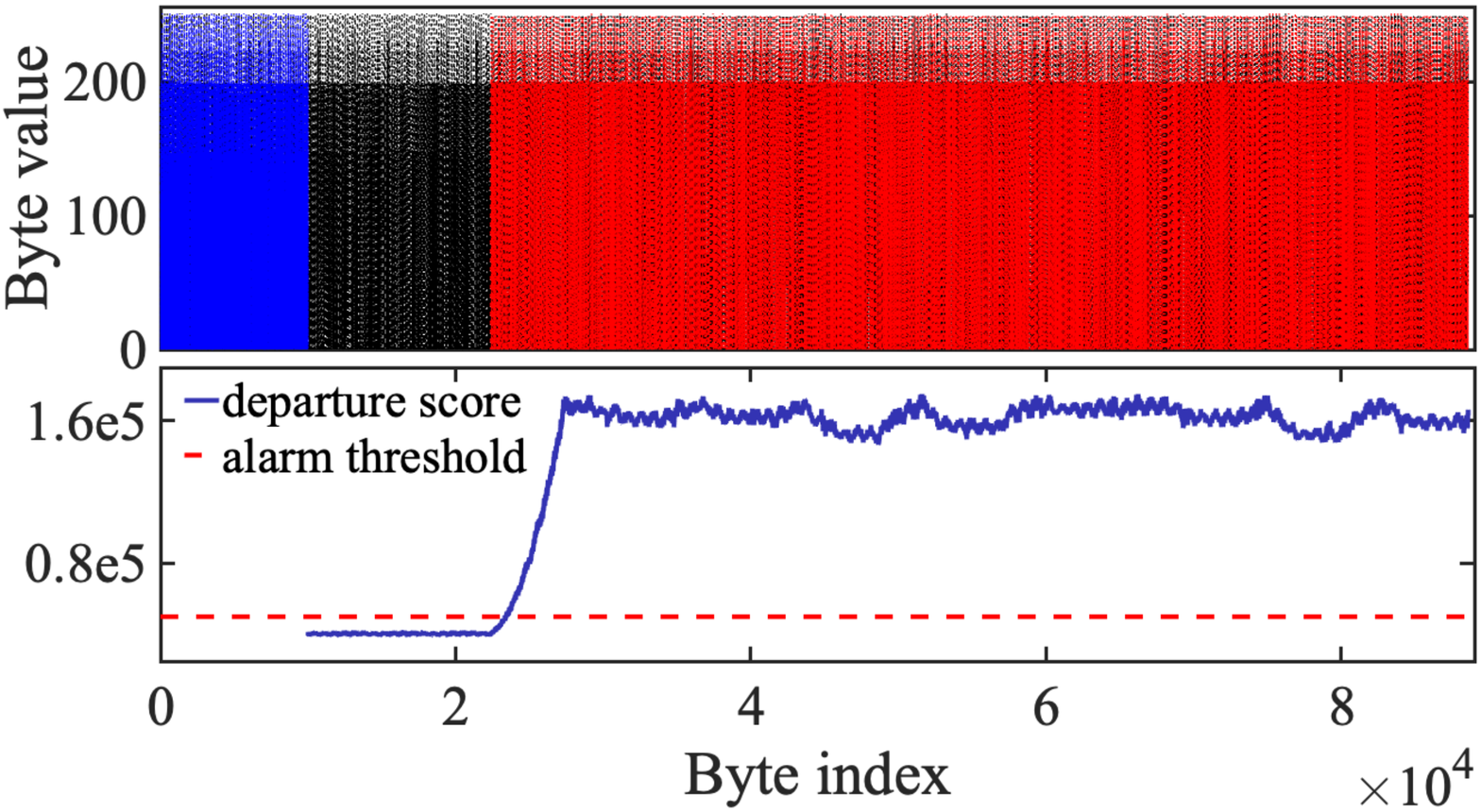}
        \caption{Fabrication on prototype}
        \label{fig:fabrication_attack_prototype}
    \end{subfigure}
    \hfill
    \begin{subfigure}[t]{0.49\columnwidth}
        \centering
        \includegraphics[width=\linewidth]{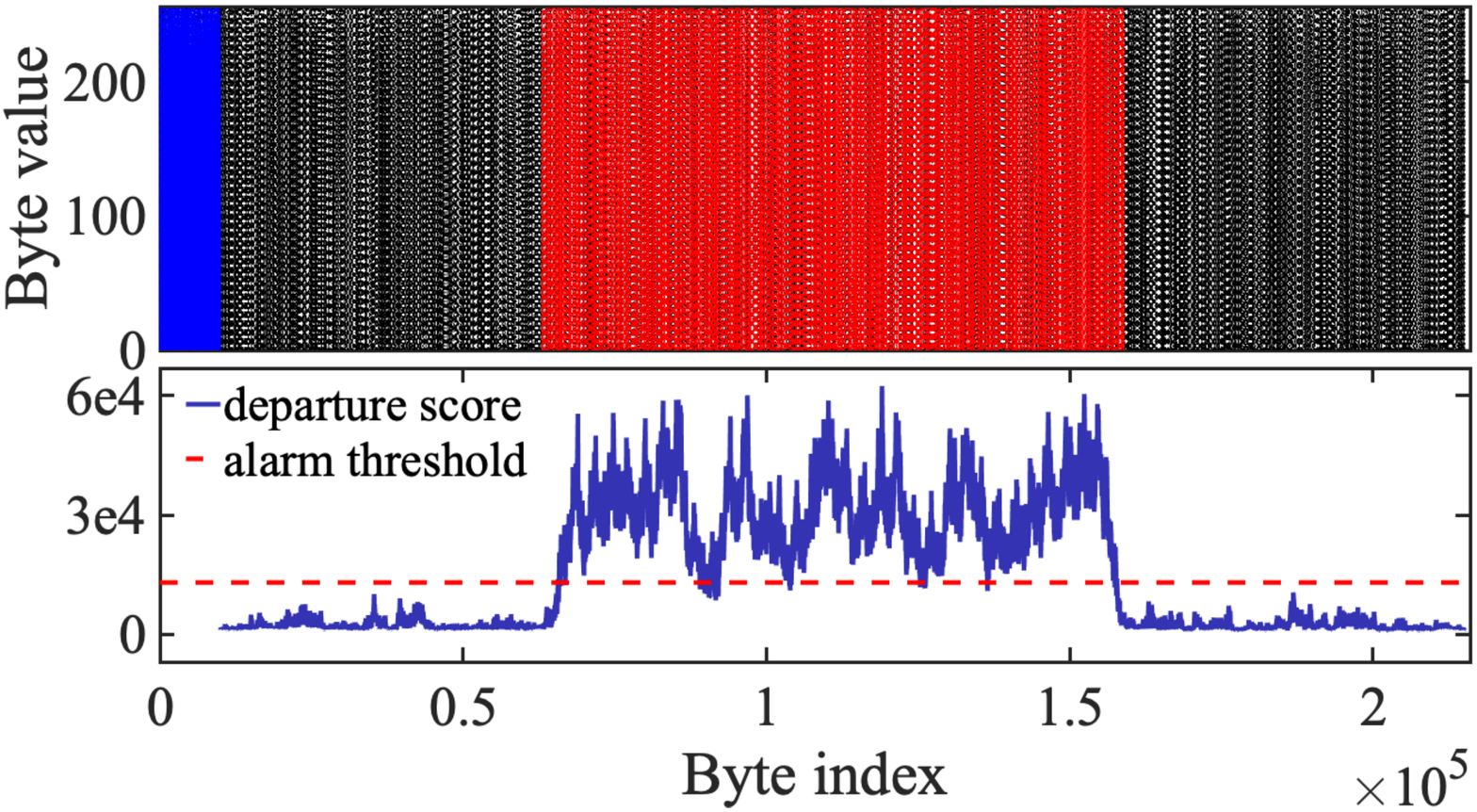}
        \caption{Fabrication on Volvo XC60}
        \label{fig:fabrication_attack_xc60}
    \end{subfigure}
    \vfill
    \begin{subfigure}[t]{0.49\columnwidth}
        \centering
        \includegraphics[width=\linewidth]{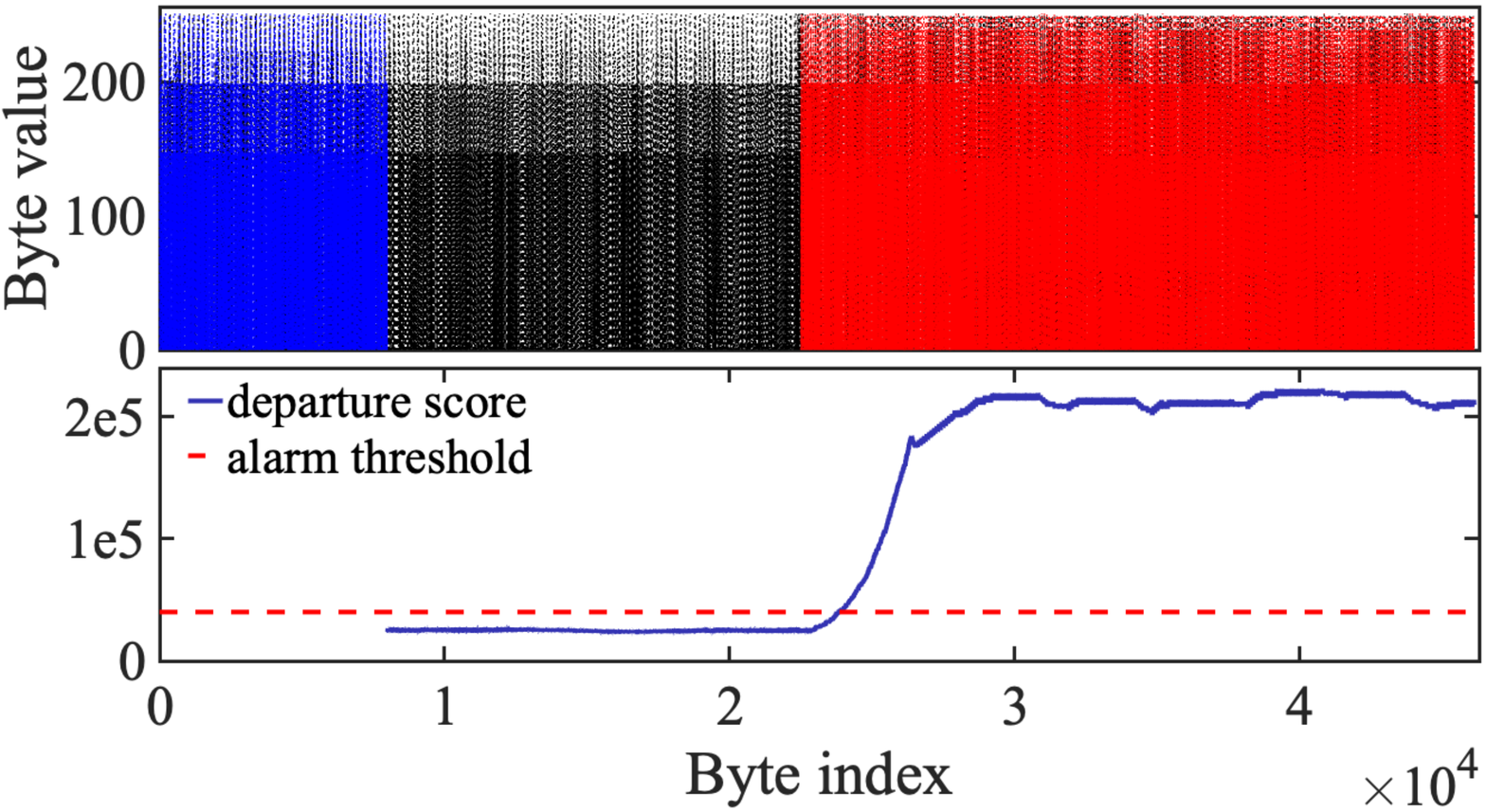}
        \caption{Masquerade on prototype}
        \label{fig:masquerade_attack_prototype}
    \end{subfigure}
    \hfill
    \begin{subfigure}[t]{0.49\columnwidth}
        \centering
        \includegraphics[width=\linewidth]{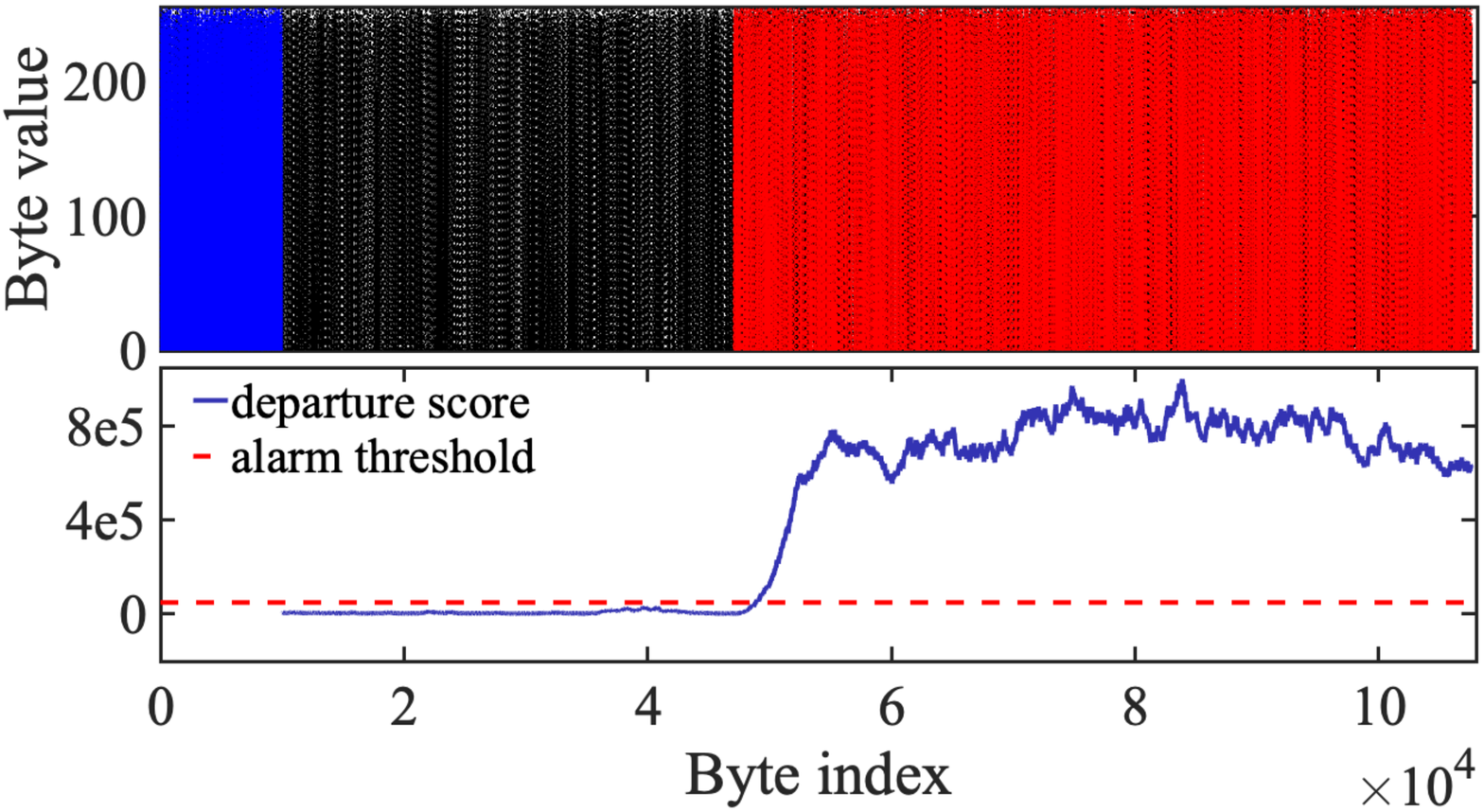}
        \caption{Masquerade on Volvo XC60}
        \label{fig:masquerade_attack_xc60}
    \end{subfigure}
    \caption{Detection of attacks on prototype and Volvo XC60.}
    \label{fig:attacks_xc60}
\end{figure}
% \caption{Detection of suspension, fabrication, and masquerade attacks on the CAN bus prototype.}
%    \label{fig:attacks_prototype}
%\end{figure}

%\begin{figure*}[!ht]
%    \centering
    
\noindent\textit{\textbf{Real vehicle.}}
In order to evaluate \acro{} in a real-world setting and demonstrate its capability of detecting attacks on recent vehicles, we use a brand new 2018 Volvo XC60 (Figure~\ref{fig:xc60}). Being a test vehicle, provided by Volvo Cars for the purpose of this research, it comes equipped with extra network interfaces, which granted us the privilege of accessing ECUs that are otherwise unreachable through the standard interfaces on a released vehicle. This privilege enabled us to perform the attacks in a much more cost-effective and time-efficient manner. It should be noted that mounting some of the attacks without such privileges requires extensive technical knowledge and would be highly costly and challenging.%, if not infeasible. [To return in the CR version]

To communicate with the vehicle's internal network, we chose Vector CANoe, a widely used software for ECU development and testing in the automotive industry, running on a laptop, together with a Vector VN1630A CAN interface connected to the vehicle's OBD-II port (see Figure~\ref{fig:in-car-obd}). This setup permitted us to create two virtual ECUs, namely $\mathbb{E}_1$ and $\mathbb{E}_2$, that were used for running \acro{} and mounting attacks respectively. 
%and a 2018 Volvo XC40 (Figure~\ref{fig:xc40})
%. The vehicle's OBD-II port (Figure~\ref{fig:in-car-obd}) was used to connect a Vector VN7600 CAN interface 
%nodes in our CAN bus prototype 
To evaluate the performance of \acro{} under complex conditions in a real vehicle, we chose to monitor one of the CAN buses on the XC60 that reflects the vehicle dynamics. The highly loaded CAN bus that we monitored with \acro{} consists of more than five safety-critical ECUs and a gateway that enables cross-domain communication. Such an environment would be a highly attractive target for attackers as it controls some major functions in the vehicle and contains messages with high integrity requirements. 
%in ters well as more than five safety critical ECUs among others which represents a  in-vehicle CAN bus
 %and . one of the most These node were connected to a CAN bus on the XC60 consisted of a gateway and more than five safety critical ECUs among others. 
There were more than 80 CAN messages transmitted at high frequencies between the ECUs on the bus, carrying approximately 1100 distinct signals in total. Thus, the chosen bus represents a difficult case for detecting IVN attacks considering the scale and complexity of the traffic. 
%In fact, as we show later, the traffic size in our experiments is approximately $38\%$ and $73\%$ more than 
In this setup, $\mathbb{E}_1$ and $\mathbb{E}_2$ were practically considered as two new nodes added to the CAN bus capable of both passively monitoring the traffic in real-time and actively influencing the vehicle's internal communication by injecting crafted messages. 
%the node $\mathbb{E}_1$ was used to perform the attacks while $\mathbb{E}_2$ was running \acro{}.
%while one of the other nodes mounted attacks. 
The evaluation was performed while the vehicle was driven on a test track. However, for safety reasons, the vehicle was in parking mode when mounting an attack.
%on the in-vehicle network.

\noindent\textit{\textbf{Captured CAN traffic.}}
To investigate the performance of \acro{} in the absence of attacks, in addition to testing the method on traffic generated by the CAN bus prototype and the real vehicle, we used CAN traffic captured from a 2012 Toyota Corolla by~\citet{mueller2012assessing} and from a 2012 Honda Civic by~\citet{diacon2013accuracy}. The Toyota Corolla traffic consists of 58 distinct CAN messages periodically transmitted over different frequencies ranging from 9\textit{ms} to 1.06 seconds. The Honda Civic traffic consists of 46 distinct CAN messages periodically transmitted over different frequencies ranging from 9\textit{ms} to 302\textit{ms}. In both cases, the data was captured while the subject vehicle was being driven under attack-free conditions. 
%Unlike the 2018 Volvo XC60, which we had access to its proprietary signal database to verify our results, we did not have access to the underlying signal specification of the vehicles from which the data was captured. Such access, however, was hardly needed because the purpose of using the captured traffic was to validate our approach under attack-free conditions.
%\mynote{CAN log from Toyota Corolla 2012. This is to show \acro{} is independent of the underlying IVN traffic. However, we only have the normal traffic in this case. I think it is worth to show that \acro{} generates no false negative/positives when used on traffic from another brand.}

%\\We need to explain why we chose this resolution}

\subsection{Results Overview}
\label{subsec:resultsOverview}
In all subsequent figures, the upper subplot displays the time series of raw CAN bytes. The part highlighted in blue corresponds to the subseries that was used for training \acro{}, and the part highlighted in red corresponds to the time frame during which the attack was occurring. Figures~\ref{fig:attacks_xc60} and~\ref{fig:evaluation-direct} display the detection results for all four attacks. Evidently, \acro{} successfully detects misbehaviors in the CAN traffic when the IVN is subject to all of the discussed attacks, including the stealthy conquest attack. Figure~\ref{fig:evaluation-normal} shows how, with a proper choice of alarm threshold, \acro{} triggers no false alarms during attack-free monitoring of CAN traffic generated from various sources. Next, we describe the experiments we performed. For each one of the four attack scenarios discussed in Section~\ref{sec:attackTaxonomy}, we performed the attack on the CAN bus prototype (Figure~\ref{fig:can-bus-prototype}) and on a 2018 Volvo XC60 (Figure~\ref{fig:xc60}).

\subsection{The Suspension Attack Experiment}
\label{subsec:detectionOfSuspensionAttack}
To simulate a suspension attack on the \emph{CAN bus prototype}, we programmed ECU $\mathbb{B}$ to stop transmitting all of its messages, namely the ones with IDs 0x01 and 0x05, 20 seconds after the start of the experiment. To perform the attack on the \emph{real vehicle}, we used ECU $\mathbb{E}_1$ to monitor the traffic on the CAN bus, while an additional OBD-II interface in the test vehicle was used to suspend a real (target) ECU on the bus by putting it into programming mode 20 seconds after establishing a diagnostic session. To observe how traffic behaves before and after a suspension attack, we put the target ECU back online on the bus by terminating the programming session.

As explained in Section~\ref{sec:methodology}, \acro{} monitors the CAN traffic at the payload level by processing one byte of traffic at a time. Although in a suspension attack, the adversary only suspends the target ECU and does not manipulate the payloads of the CAN messages, \acro{} still manages to detect the attack as indicated in Figures~\ref{fig:suspension_attack_prototype} and~\ref{fig:suspension_attack_xc60}. The reason why this attack is detectable by \acro{} is that the suspension of the target ECU leads to a change in the byte sequence such that the subsequences monitored during the attack would be foreign to \acro{} because they were unseen during the learning phase. This observation effectively implies that any change in the frequency of transmission should be detectable by \acro{} because it would induce a change in the payload byte sequence.
%To achieve this, we remove the target ECU's message IDs from $\mathbb{P}$'s acceptance filter list at around 20 seconds into the experiment, thus emulating a suspension attack. 

\subsection{The Fabrication \& Masquerade Attacks}
\label{subsec:detectionOfFabricationAttack}
\textit{\textbf{Fabrication Attacks}.} To evaluate \acro{} against fabrication attacks on the \emph{CAN bus prototype}, we programmed ECU $\mathbb{A}$ to start injecting messages with ID 0x05 at the same frequency as its original transmitter $\mathbb{B}$ roughly 20 seconds after the start of the experiment. The payload of the injected message is identical to the original data sent by $\mathbb{B}$ except for the last byte, which we assumed to represent the target signal for the adversary. Therefore, $\mathbb{A}$ was programmed to change only the last byte of the payload when injecting forged messages with ID 0x05. To carry out the attack on the \emph{real vehicle}, we considered a scenario in which ECU $\mathbb{E}_2$ is fully compromised by the adversary and used to mount a fabrication attack on a real ECU in the vehicle. Using the proprietary signal database, we identified the ECU that transmits Speed Limit Warning (SLW) signals in a CAN message to the gateway ECU, which subsequently forwards it to the dashboard ECU located in another network domain. Ignoring the various signals that were irrelevant to the SLW function, we located the bits in the message payload that belong to the SLW signal; specifically, the bits which trigger the warning and specify whether it shall be visual, audible, or both. Finally, we programmed $\mathbb{E}_2$ to initiate a fabrication attack on the SLW message.

\textit{\textbf{Masquerade Attacks.}} To launch a masquerade attack on the \emph{CAN bus prototype}, we programmed ECU $\mathbb{A}$, assumed to be fully compromised, to launch an attack on ECU $\mathbb{B}$, assumed to be partially compromised, in which $\mathbb{B}$ is forcibly withdrawn from transmission after 20 seconds, and immediately afterwards, $\mathbb{A}$ starts to inject messages with ID 0x05 at the original frequency on behalf of $\mathbb{B}$, but with a maliciously crafted payload. To perform a masquerade attack on the \emph{real vehicle}, we considered a scenario in which ECU $\mathbb{E}_2$, assumed to be fully compromised by the adversary, launches a fabrication attack on the real ECU ($\mathbb{T}$) in the vehicle that is responsible for sending the engine rotation speed (also known as Revolutions Per Minute (RPM)) values to the dashboard ECU and a few other receivers. For this attack scenario, we used the signal database to identify the ECU responsible for transmitting CAN messages containing RPM values (among other sensor values) to the gateway ECU, which in turn relays them to the dashboard ECU located in another network domain. We located the bits in the message payload that belong to the RPM signal and programmed $\mathbb{E}_2$ to transmit the maliciously altered RPM message, while immediately suspending $\mathbb{T}$. This was achieved by injecting forged RPM messages from $\mathbb{E}_2$ having the same ID and frequency as the original RPM message, but with forged RPM values in the payload, while placing $\mathbb{T}$ into a suspension mode by sending a reprogramming request through a diagnostic session.

In the fabrication attack scenario, the payloads corresponding to SLW messages are maliciously altered and, as in the suspension attack, the frequency of messages is affected as well, since ECU $\mathbb{A}$ transmits the fabricated messages at a higher frequency. As pointed out in Section~\ref{subsec:attackScenarios}, when a fabrication attack is not feasible due to the receiving ECU having protection mechanisms in place, a \emph{masquerade attack} can be performed instead. Technically, the masquerade attack causes similar changes in the payload byte sequence, since it involves both altering the payload of the intended message and suspending the target ECU. Figures~\ref{fig:fabrication_attack_prototype},~\ref{fig:masquerade_attack_prototype},~\ref{fig:fabrication_attack_xc60} and~\ref{fig:masquerade_attack_xc60} clearly show how both attacks can be detected by \acro{}.
%, resulting in the dashboard continuously exhibiting false visual and audible warnings, which may distract the driver and possibly lead to an accident

\subsection{The Conquest Attack Experiment}
\label{subsec:detectionOfDirectAttack}
To simulate a stealthy conquest attack on the \emph{CAN bus prototype}, we assumed that the target ECU $\mathbb{B}$ is fully compromised by an adversary who wants to silently manipulate specific signals in one or several messages without injecting any other messages into the bus by any other ECUs. To perform this attack, $\mathbb{B}$ was programmed to maliciously modify only the last two bytes in the payload of messages with ID 0x05 in such a way that the modified values remain within the normal range of the corresponding signal.

\begin{figure}[!t]
    \centering
    \begin{subfigure}[t]{0.49\columnwidth}
        \centering
        \includegraphics[width=\linewidth]{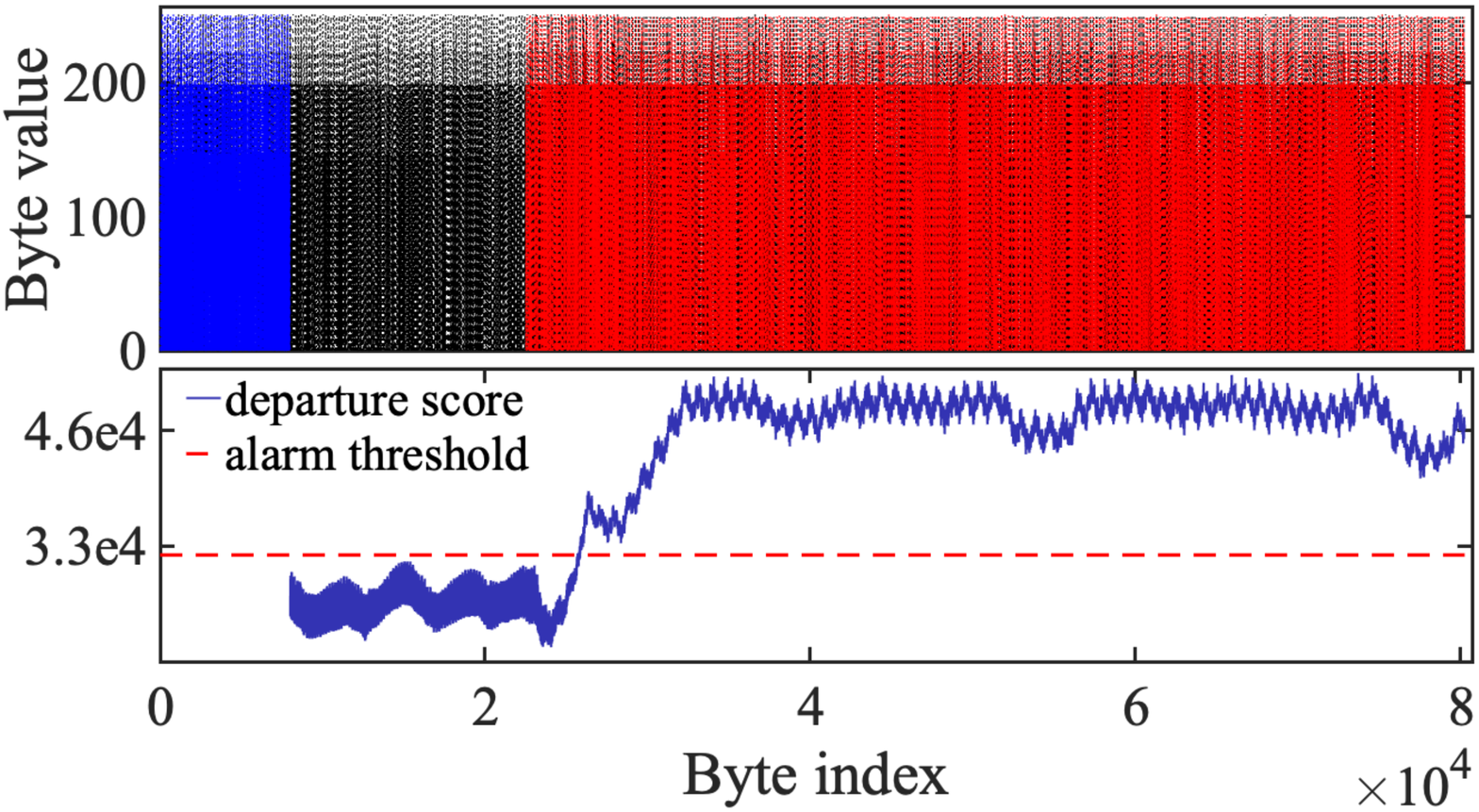}
        \caption{On CAN bus prototype}
        \label{fig:direct_attack_prototype}
    \end{subfigure}
    \hfill
    \begin{subfigure}[t]{0.49\columnwidth}
        \centering
        \includegraphics[width=\linewidth]{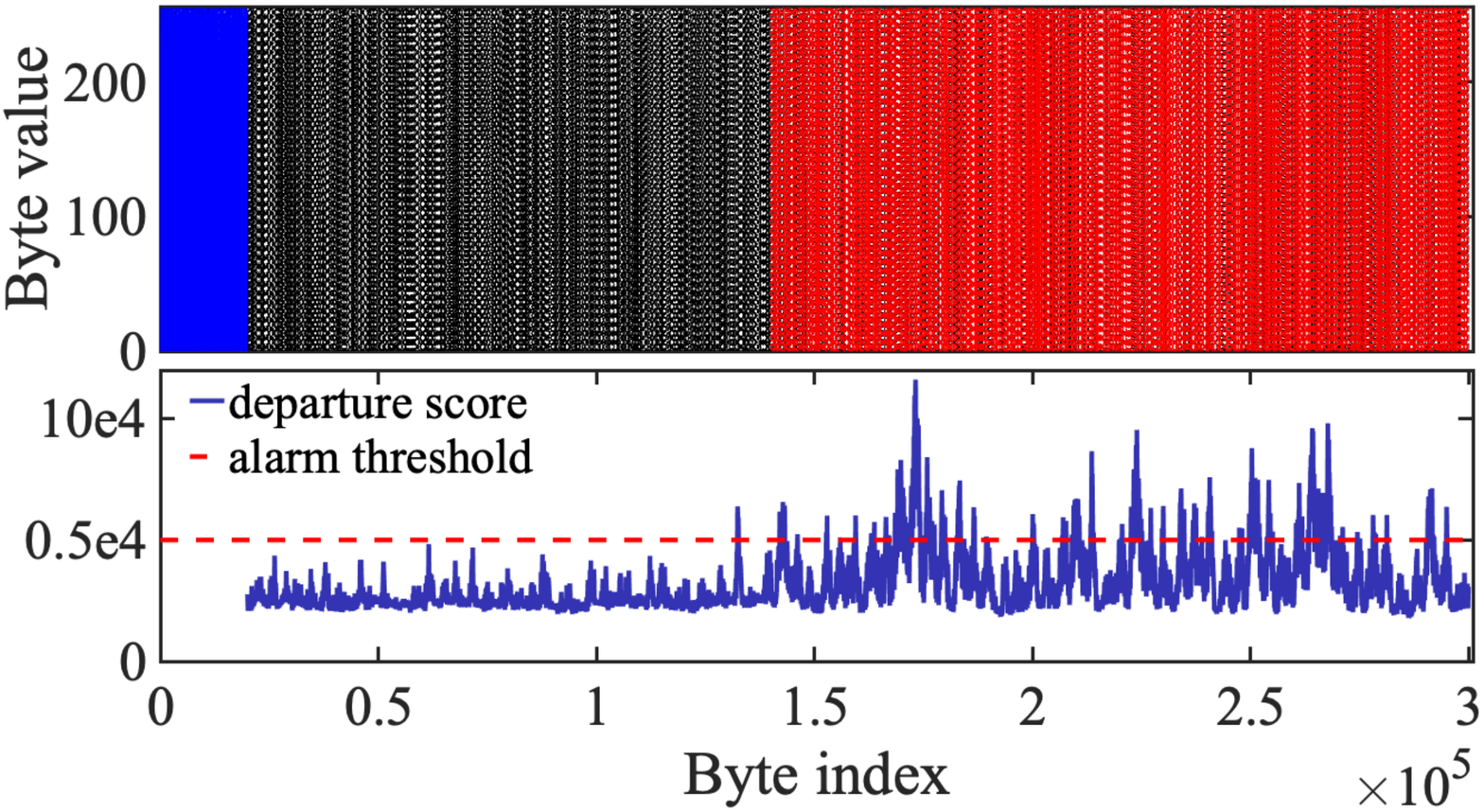}
        \caption{On 2018 Volvo XC60}
        \label{fig:direct_attack_xc60}
    \end{subfigure}
    \caption{Detection of the stealthy conquest attack.}
    \label{fig:evaluation-direct}
\end{figure}

%For reasons related to the complexity of reprogramming a real ECU software (knowledge, time, and possibility of breaking an ECU), it would not have been reasonable to perform the conquest attack on the real vehicle. Instead, we chose to emulate the attack on the Volvo XC60 according to the following setup. We assumed that the target ECU ($\mathbb{E}_2$) is fully compromised and that the attacker is capable of changing certain bytes of the payload where one or more critical signals are stored, while leaving the remaining data bytes intact. First, we created a new message (ID 0x7EA) such that it has the lowest priority on the bus so as to avoid causing any conflicts in the bus arbitration phase. Then, we programmed $\mathbb{E}_2$ to transmit this message for the duration of 20 seconds, and immediately afterwards, we altered the last two bytes of the payload and continued the transmission. For the sake of stealthiness, we made sure that the altered values already existed in the traffic and were repeated often enough. 
To perform the conquest attack on the \emph{real vehicle}, we considered a scenario in which the target ECU $\mathbb{T}$ is fully compromised and the adversary is capable of changing certain bytes of the payload where one or more critical signals are stored, while leaving the remaining data bytes intact. In this scenario, an existing safety-critical diagnostic service (see Section~\ref{subsec:can}) used for overwriting specific signal values on the target ECU is identified as the launching point of the attack. As this service is protected from unauthorized users by the Security Access request-response protocol, an additional OBD-II interface in the test vehicle would run the Security Access algorithm and unlock the service on the target ECU, while ECU $\mathbb{E}_1$ monitors the traffic on the CAN bus. In this attack, the signal related to fuel consumption was maliciously altered roughly 20 seconds after the start of the experiment. To ensure stealthiness, we made sure that the altered values already existed in the traffic and were repeated often enough.
%and was meant to be used during the test and development phase. 

Unlike existing attacks, in a conquest attack, only a relatively small subset of payload bytes is affected, while all messages are transmitted by their original ECUs. Importantly, the adversary would inject the attack message using its \emph{original sender} ECU and at the \emph{original frequency}. This effectively means that the adversary is able to evade protection mechanisms, and both timing-based and fingerprinting techniques discussed in Section~\ref{subsec:relatedwork}, since neither the timing behavior nor the low-level physical specifications of the messages are violated. In contrast to the other attacks, the conquest attack causes minimal changes in the CAN traffic. Nonetheless, as \acro{} is inherently capable of detecting slight variations in the monitored traffic, as shown in Figure~\ref{fig:evaluation-direct}, it successfully detects the conquest attack both on the CAN-bus prototype and the real vehicle. The significance of this result particularly lies in \acro{}'s distinctive capability of detecting subtle anomalous deviations in a small fraction of signal data bytes on a heavily loaded CAN bus containing more than 1100 distinct signals, and without prior knowledge of the underlying signal specifications. 
\begin{figure}[!b]
    \centering
    \begin{subfigure}[t]{0.49\columnwidth}
        \centering
        \includegraphics[width=\columnwidth]{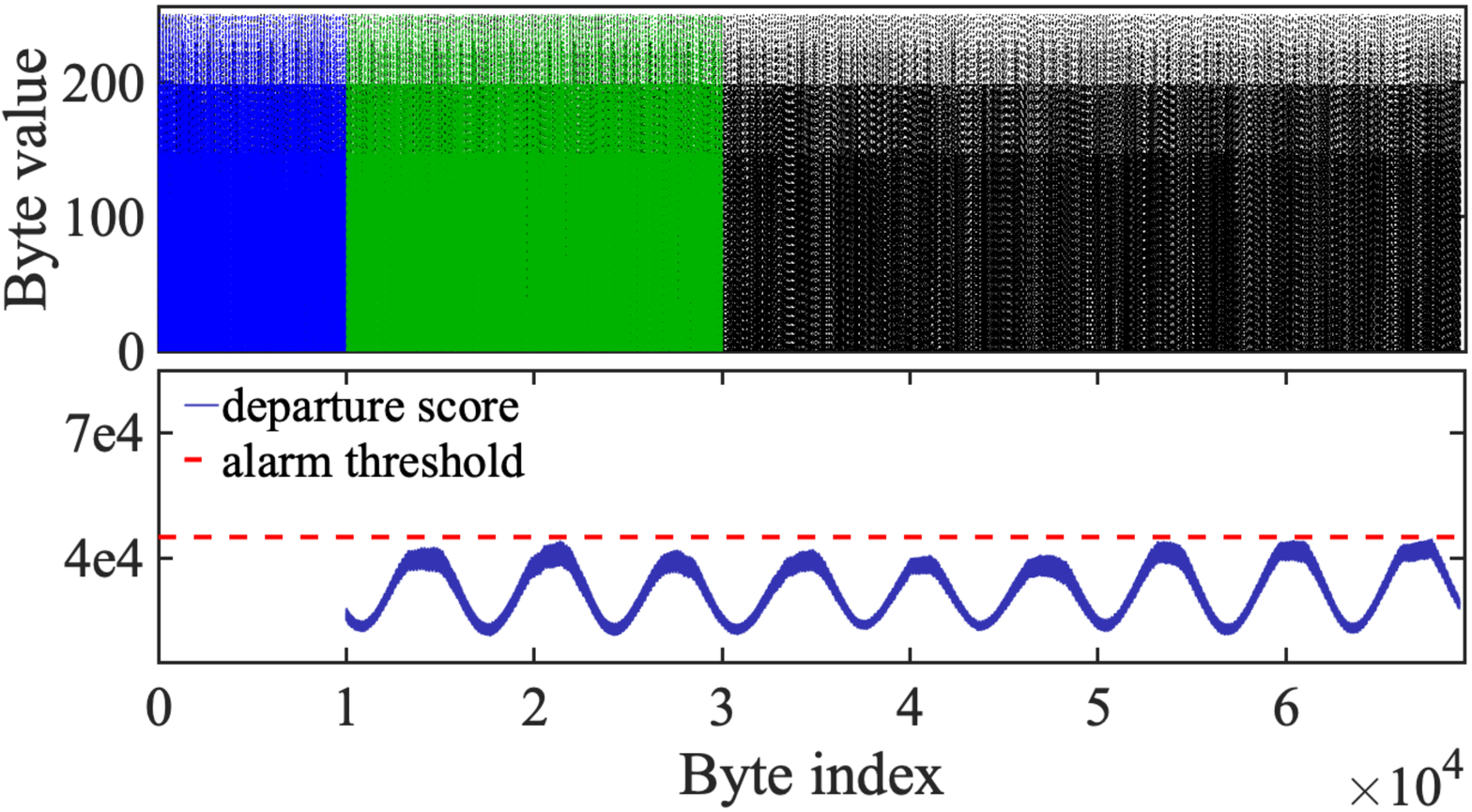}
        \caption{CAN bus prototype}
        \label{fig:normal-can-setup}
    \end{subfigure}%
    \hfill
    \begin{subfigure}[t]{.49\columnwidth}
        \centering
        \includegraphics[width=\columnwidth]{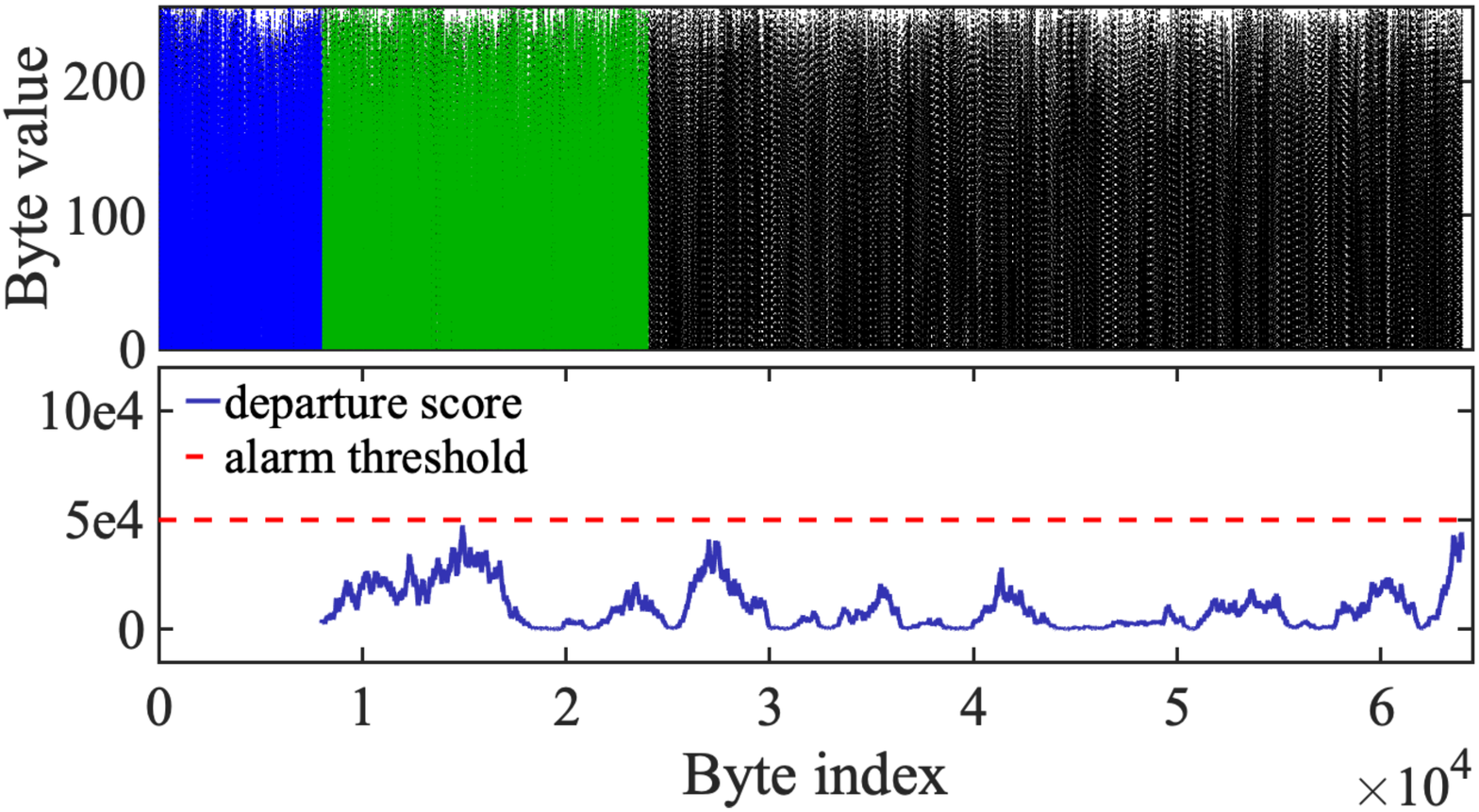}
        \caption{2018 Volvo XC60}
        \label{fig:normal-xc60}
    \end{subfigure}%
    \vfill
    \begin{subfigure}[t]{0.49\columnwidth}
        \centering
        \includegraphics[width=\columnwidth]{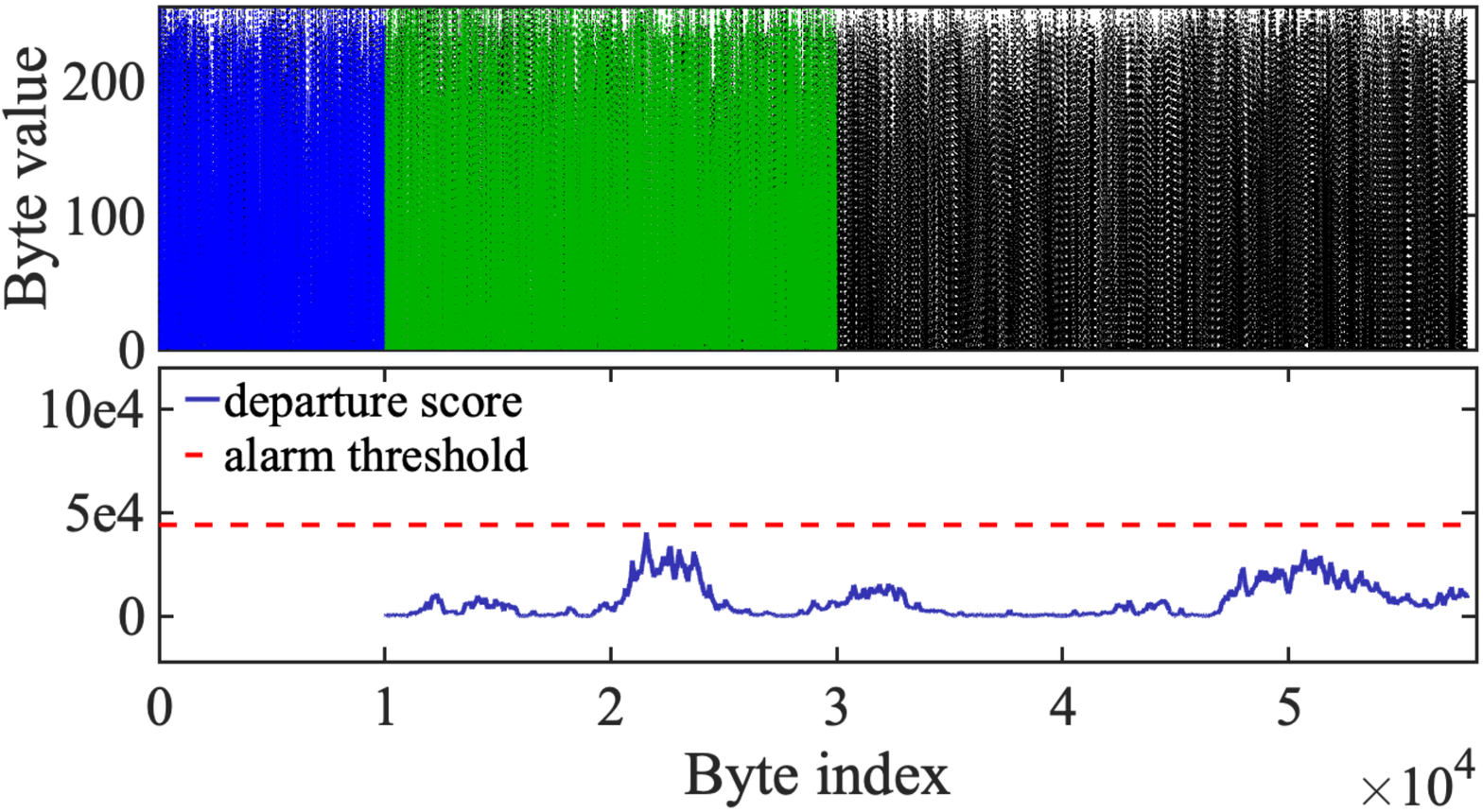}
        \caption{2012 Toyota Corolla}
        \label{fig:toyota-corolla-normal}
    \end{subfigure}%
    \hfill
    \centering
    \begin{subfigure}[t]{0.49\columnwidth}
        \centering
        \includegraphics[width=\columnwidth]{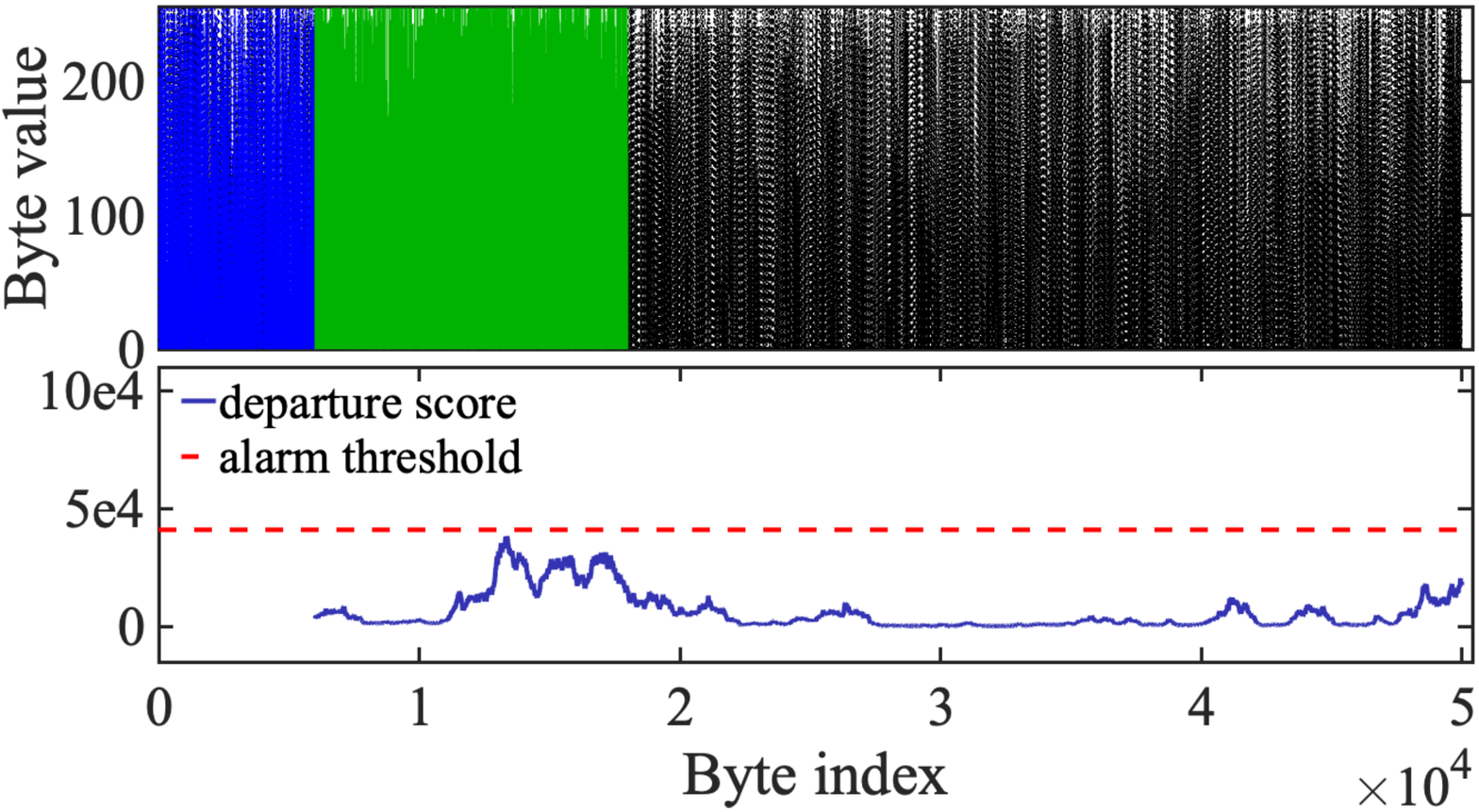}
        \caption{2012 Honda Civic}
        \label{fig:honda-civic}
    \end{subfigure}%
    \caption{Evaluation of \acro{} on attack-free CAN traffic.}
    \label{fig:evaluation-normal}
\end{figure}

\begin{figure}[!t]
\centering
\begin{subfigure}[t]{\columnwidth}
\includegraphics[width=\textwidth]{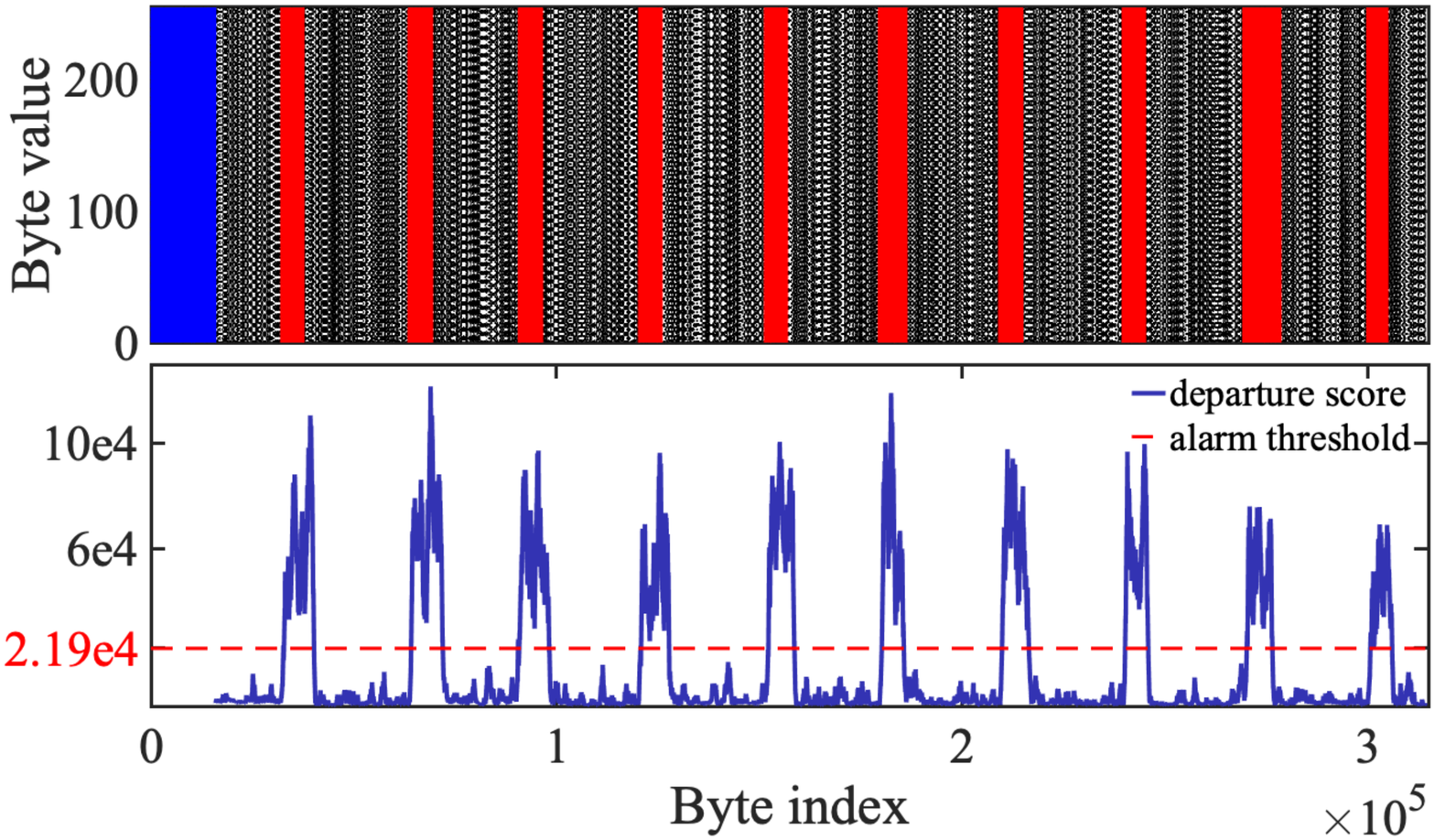}
\caption{Repeated masquerade attacks}
\label{fig:masq_10}
\end{subfigure}%
\vfill
\begin{subfigure}[t]{\columnwidth}
\centering
\includegraphics[width=\textwidth]{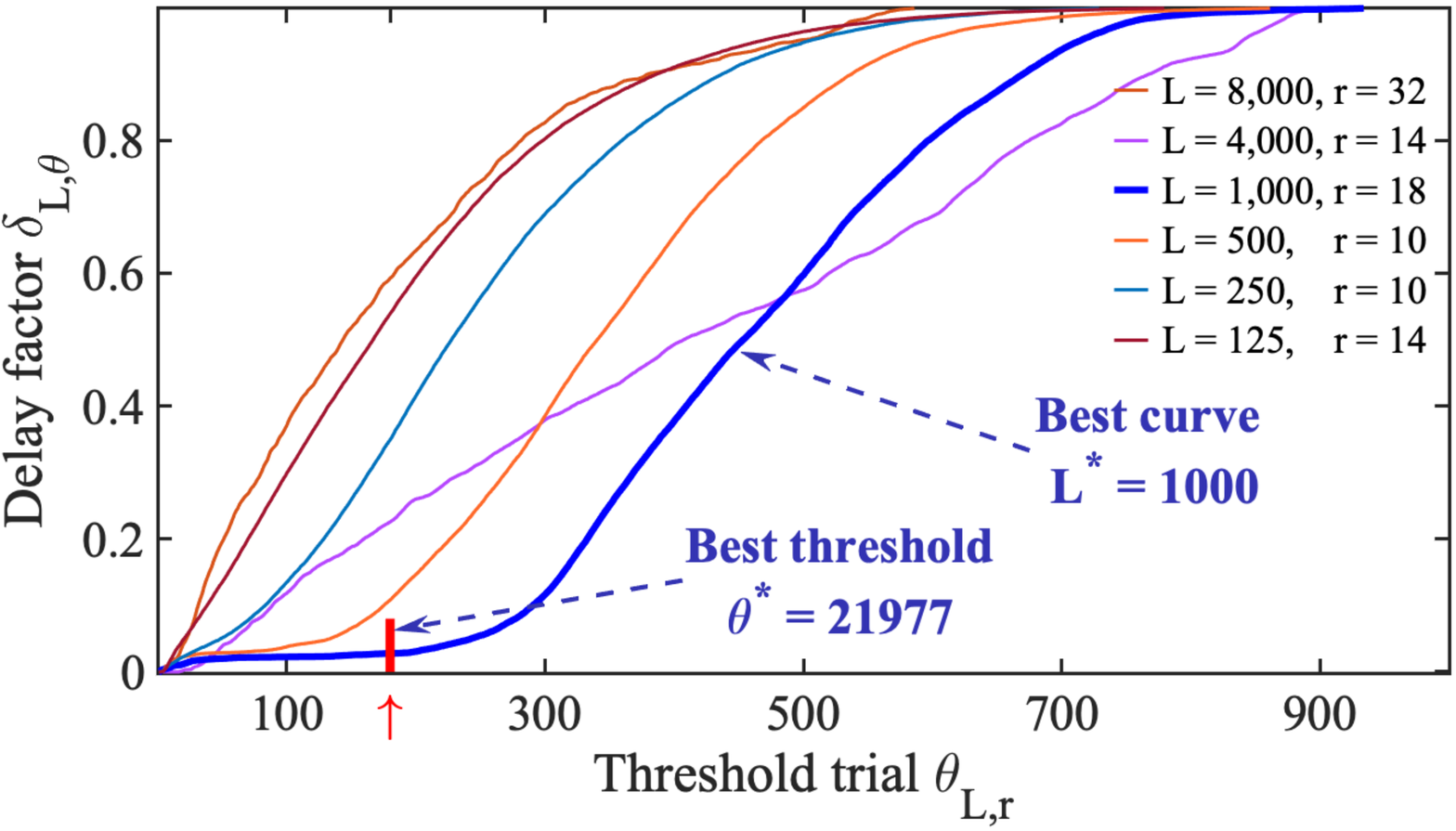}
\caption{Identifying the best threshold}
\label{fig:best_thresh}
\end{subfigure}%
\caption{Optimizing the detection delay/false alarms trade-off.}
\end{figure}
\subsection{The Attack-Free Experiment}
\label{subsec:normalTrafficEvaluation}
To simulate normal traffic behavior on the CAN bus prototype, we programmed ECUs $\mathbb{A}$ and $\mathbb{B}$ to transmit messages 0x1C and [0x01,0x05] respectively. For the sake of producing a realistic behavior, the payloads of the messages were constructed to be similar to that of existing messages on the Volvo XC60 CAN bus. In order to investigate the performance of \acro{} under normal driving conditions, the test vehicle was driven in an urban area, where unforeseen events and ambient conditions required the driver to show different reactions (sudden acceleration, braking, etc.), thus adding normal, yet significant, variability to the IVN traffic, which was passively monitored by $\mathbb{E}_1$ throughout the experiment. Finally, in order to demonstrate the applicability of our approach to other vehicle brands, we ran \acro{} on normal CAN traffic captured from a 2012 Toyota Corolla and a 2012 Honda Civic.
%\footnote{As a disclaimer, we wish to stress that this experiment did not involve performing any attacks, and thus there were no safety concerns.} 

Figure~\ref{fig:evaluation-normal} shows the behavior of \acro{} when performing on normal traffic generated by the CAN prototype and three different real vehicles. In particular, the CAN traffic displayed in Figure~\ref{fig:normal-xc60} was captured while driving the Volvo XC60 test vehicle in an urban area and, as indicated in the figure, no false alarms were triggered by \acro{}. In order to determine the alarm threshold for these experiments, we followed the same methodology adopted in~\cite{aoudi2018pasad}, where we ran \acro{} on a \emph{validation} subseries of CAN traffic (highlighted in green), selected to be twice as long as the training subseries, and then defined the threshold to be slightly higher than the maximum departure score attained during the validation period. A more rigorous choice of threshold is described next.

\subsection{Quantitative Analysis \& Best Threshold}
\label{subsec:choiceOfThreshold}
As demonstrated by the previous experiments, \acro{} can detect attacks on IVNs with maximum accuracy and without triggering any false alarms. Therefore, we do not evaluate \acro{}'s performance using traditional classification metrics such as ROC curves. However, one important metric for evaluating \acro{}'s performance is the average time it takes to detect an attack, which we refer to as the \emph{delay factor}. Evidently, setting the alarm threshold high enough so as to avoid false positives results in a larger delay factor. More precisely, the delay factor, denoted by $\delta_\ms{L,\theta}$, varies with the lag parameter $L$ and is inversely proportional to the alarm threshold $\theta_\ms{L,r}$ for a fixed choice of $L$ and $r$.

We define the delay factor as the number of traffic instances (bytes) that fall into the aggregate of attack intervals $T_\ms{a}$ and whose corresponding departure scores are below the threshold, normalized by the total number of attack instances $\gamma_\ms{a}$. More formally, $\delta_\ms{L,\theta} = \bigm|\{\mathcal{T}_\ms{i}\ |\ i\in T_\ms{a},\ \tilde{D}_\ms{i} < \theta\}\bigm|\ \bigm/ \gamma_\ms{a}$.

Figure~\ref{fig:masq_10} shows CAN traffic from an IVN that was subject to 10 repeated masquerade attacks. To discover the \emph{best} alarm threshold, for $k$ different values of the lag parameter $L$, we determine the corresponding statistical dimension $r$, and compute $\delta_\ms{L,\theta}$ for $1,000$ different values of $\theta$ selected by evenly dividing the range between the minimum and maximum departure scores attained throughout the entire experiment. As illustrated in Figure~\ref{fig:best_thresh}, this procedure yields $k$ curves describing the trade-off between the time to detection and the likelihood of false positives for different choices of \acro{}'s lag parameter. The task then is to optimize this trade-off by determining the best delay factor $\delta_\ms{L^*,\theta^*}$. Intuitively, the best lag parameter $L^*$ should correspond to the curve with the minimum AUC, which measures the total detection delay over the different threshold trials. After identifying the best curve as such, we determine the best threshold $\theta^*$ by visually identifying the optimal \emph{cut} corresponding to the threshold that minimizes the delay factor while maximizing the threshold, where the latter is equivalent to minimizing the likelihood of false positives. The approach to determining the alarm threshold mentioned in Section~\ref{subsec:normalTrafficEvaluation} is naive and not very tolerant to changes in system dynamics. By contrast, as shown in Figure~\ref{fig:masq_10}, the described procedure results in a more sensible choice of threshold that decreases the sensitivity to false alarms while maintaining prompt detection of attacks. 

\section{Conclusion}
\label{sec:conclusion}
Over the past decade, IVNs have witnessed a rapid increase in the number of cyberattacks, raising concerns about the safety of passengers. With CAN being the most prevalent protocol used for safety-critical applications in vehicles, designing intrusion detection systems for CAN communication has become a major area of interest. 
%Despite being capable of detecting attacks that cause obvious deviations in the highly regular properties of IVN communications, the proposed methods in literature suffer from two major limitations: they are dependent on the underlying system specifications and they are incapable of detecting critical stealthy attacks. 
This paper has made several noteworthy contributions to the field of automotive security. First, we have presented \acro{}, an efficient attack-detection mechanism that is particularly suitable for the IVN domain. Second, we have introduced the truly stealthy conquest attack with potential to cause serious impact on vehicles. Third, we have demonstrated, through extensive experiments including performing attacks on a 2018 Volvo XC60 test vehicle, how unlike existing methods, \acro{} is capable of detecting stealthy attacks on IVNs. Finally, we have shown that \acro{} enjoys the advantage of being specification-agnostic, which makes it applicable to a wide range of vehicle models and deployable in real-world settings.

\footnotesize
\bibliographystyle{IEEEtranN}
\bibliography{references.bib}

\end{document}